\documentclass[12pt]{article}

\usepackage{amsmath}
\usepackage{amssymb}
\usepackage{epsfig}
\usepackage{graphicx}
\usepackage{cite}
\usepackage{multirow}
\usepackage{longtable}
\usepackage{lscape}
\usepackage{dcolumn}
\usepackage[dvips]{color}
\usepackage{rotating}

\allowdisplaybreaks[4] % for page break

\newcommand{\capdef}{}
\newcommand{\mycaption}[2][\capdef]{\renewcommand{\capdef}{#2}%
        \caption[#1]{{\footnotesize #2}}}
\makeatletter
\renewcommand{\fnum@table}{\textbf{\tablename~\thetable}}
\renewcommand{\fnum@figure}{\textbf{\figurename~\thefigure}}
\makeatother

\newcounter{myenumi}

\renewcommand{\themyenumi}{\roman{myenumi}}
{\end{list}}

\newlength{\myem}
\newcounter{mysubequation}[equation]

\makeatletter
\renewcommand{\section}{\@startsection{section}{1}{0em}{-\baselineskip}%
{\baselineskip}{\normalfont\large\bfseries}}
\renewcommand{\subsection}%
{\@startsection{subsection}{2}{0em}{-0.7\baselineskip}%
{0.7\baselineskip}{\normalfont\bfseries}}
\makeatother

\newcommand{\bi}{\begin{itemize}}
\newcommand{\ei}{\end{itemize}}

\newcommand{\be}{\begin{equation}}
\newcommand{\ee}{\end{equation}}
\newcommand{\bea}{\begin{eqnarray}}
\newcommand{\eea}{\end{eqnarray}}
\newcommand{\nn}{\nonumber}
\newcommand{\ldm}{\Delta m_{31}^2}
\newcommand{\sdm}{\Delta m_{21}^2}
\newcommand{\vldm}{\Delta m_{41}^2}
\newcommand{\deltacp}{\delta_{\mathrm{CP}}}
\newcommand{\stheta}{\sin^2 2 \theta_{13}}

\newcommand{\ie}{{\it i.e.}}

\newcommand{\eg}{{\it e.g.}}

\newcommand{\cf}{{\it cf.}}

\newcommand{\eq}{Eq.}

\newcommand{\fig}{Fig.}

\newcommand{\Ref}{Ref.}
\newcommand{\Refs}{Refs.}
\newcommand{\Sec}{Sec.}

\newcommand{\App}{the Appendix}

\newcommand{\Tab}{Table}

\newcommand{\equ}[1]{\eq~(\ref{equ:#1})}
\newcommand{\figu}[1]{\fig~\ref{fig:#1}}

\def \mr{\mathrm}

\begin{document}
%%%%%%%%%%%%%%%%%%%%%%%%%%%%%%%%%%%%%%%%%%%%%%%%%%%%%%%%%%%%%%%%%%%%%
%%%%                     Title-page                              %%%%
%%%%%%%%%%%%%%%%%%%%%%%%%%%%%%%%%%%%%%%%%%%%%%%%%%%%%%%%%%%%%%%%%%%%%

\begin{titlepage}

\renewcommand{\thefootnote}{\alph{footnote}}

\vspace*{-3.cm}
\begin{flushright}
EURONU-WP6-10-23 \\
IDS-NF-018
\end{flushright}

%\vspace*{0.5cm}

\renewcommand{\thefootnote}{\fnsymbol{footnote}}
\setcounter{footnote}{-1}

{\begin{center}
{\large\bf
Sterile neutrinos beyond LSND at the Neutrino Factory
} 
\end{center}}

\renewcommand{\thefootnote}{\alph{footnote}}

\vspace*{.8cm}
\vspace*{.3cm}
{\begin{center} {\large{\sc
                Davide~Meloni\footnote[1]{\makebox[1.cm]{Email:}
                davide.meloni@physik.uni-wuerzburg.de}, 
                Jian~Tang\footnote[2]{\makebox[1.cm]{Email:}
                jtang@physik.uni-wuerzburg.de}, and
                Walter~Winter\footnote[3]{\makebox[1.cm]{Email:}
                winter@physik.uni-wuerzburg.de}
                }}
\end{center}}
\vspace*{0cm}
{\it
\begin{center}

\footnotemark[1]$^,$\footnotemark[2]$^,$\footnotemark[3]
       Institut f{\"u}r Theoretische Physik und Astrophysik, \\ Universit{\"a}t W{\"u}rzburg, 
       D--97074 W{\"u}rzburg, Germany

\end{center}}

\vspace*{1.5cm}

\begin{center}
{\Large \today}
\end{center}

{\Large \bf
\begin{center} Abstract \end{center}  }

We discuss the effects of one additional sterile neutrino at the Neutrino Factory. Compared to earlier analyses, which have been motivated by LSND results, we do not impose any constraint on the additional mass squared splitting. This means that the additional mass eigenstate could, with small mixings, be located among the known ones, as it is suggested by the recent analysis of cosmological data. We use a self-consistent framework at the Neutrino Factory without any constraints on the new parameters. We demonstrate for a combined short and long baseline setup that near detectors can provide the expected sensitivity at the LSND-motivated $\Delta m_{41}^2$-range, while some sensitivity can also be obtained in the region of the atmospheric mass splitting from the long baselines. We point out that limits on such very light sterile neutrinos may also be obtained from a re-analysis of atmospheric and solar neutrino oscillation data, as well as from supernova neutrino observations.
In the second part of the analysis, we compare our sensitivity with the existing literature using additional assumptions, such as $|\Delta m_{41}^2| \gg |\Delta m_{31}^2|$ leading to averaging of the fast oscillations in the far detectors. We demonstrate that while the Neutrino Factory has excellent sensitivity compared to existing studies using similar assumptions, one has to be very careful interpreting these results for a combined short and long baseline setup where oscillations could occur in the near detectors.
We also test the impact of additional $\nu_\tau$ detectors at the short and long baselines, and we do not find a substantial improvement of the sensitivities.

\vspace*{.5cm}

\end{titlepage}

\newpage

\renewcommand{\thefootnote}{\arabic{footnote}}
\setcounter{footnote}{0}

\section{Introduction}

Neutrino oscillation experiments recently have given us compelling evidence that active neutrinos are massive particles~\cite{GonzalezGarcia:2007ib}, pointing towards physics beyond the Standard Model. In the standard three-generation scenario, there are two characteristic mass squared splittings ($\Delta m_{31}^2\,,\Delta m_{21}^2$) and three mixing angles ($\theta_{12}\,,\theta_{13}\,,\theta_{23}$) as well as a CP violation phase $\delta_{\textrm{CP}}$ affecting neutrino oscillations. 
Disappearance of muon neutrinos, which is mainly driven by $|\Delta m_{31}^2|$ and $\theta_{23}$, has been observed in the atmospheric neutrino oscillation experiments, such as Super-Kamiokande~\cite{Fukuda:1998mi}, and in the MINOS long baseline experiment~\cite{Adamson:2008zt}. Disappearance of electron neutrinos has been observed from solar neutrino oscillation experiments very sensitive to $\theta_{12}$~\cite{Ahmad:2002jz}, whereas  $\Delta m_{12}^2$ has been strongly constrained by the KamLAND long baseline reactor neutrino experiment~\cite{Araki:2004mb}. The CHOOZ short-baseline  reactor neutrino oscillation experiment~\cite{Apollonio:2002gd} has provided a limit $\stheta \lesssim 0.1$. There are still unknown questions in the standard scenario: $\Delta m_{31}^2>0$ (normal ordering) or $\Delta m_{31}^2<0$ (inverted ordering); the value of $\theta_{13}$, as there has been a recent hint for $\theta_{13}>0$~\cite{Fogli:2008jx}, and whether there is CP violation (CPV) in the lepton sector. 
Apart from these measurements, there has been the exceptional LSND measurement with an incompatible anomaly~\cite{Aguilar:2001ty}. The simplest interpretation has been an additional  sterile neutrino added to the standard picture with $|\Delta m_{41}^2| \gg |\Delta m_{31}^2|$.  A global fit to all experimental data, however, is not in favor to this hypothesis~\cite{Maltoni:2007zf}, which means that more exotic scenarios would be required to describe this anomaly, such as a decaying sterile neutrino~\cite{PalomaresRuiz:2005vf}. The recent results from MiniBooNE, however, are consistent with sterile neutrino oscillations in the antineutrino sector~\cite{AguilarArevalo:2010wv}. Note that the LSND interpretation requires significant mixings with the active neutrinos, whereas small ad-mixtures, even if $|\Delta m_{41}^2| \gg |\Delta m_{31}^2|$, are not excluded. On the other hand, sterile neutrinos with $|\Delta m_{41}^2| \sim |\Delta m_{31}^2|$ or $|\Delta m_{41}^2| \sim \Delta m_{21}^2$, as they are motivated by a recent cosmological data analysis~\cite{Hamann:2010bk}, have been hardly studied in the literature. Of course, such sterile neutrinos have to have small mixings with the active ones in order not to spoil the leading three-flavor fits. Therefore, we discuss sterile neutrinos beyond LSND, \ie, without any constraints to $\Delta m_{41}^2$ including the full range. We consider the simplest case of only one additional sterile neutrino.  Note that we do not impose cosmological constraints on the sterile neutrino masses, but we assume that the masses are to be constrained in a self-consistent way with terrestrial experiments.

Since we cannot rely on constraints from the atmospheric and solar experiments, which need to be re-analyzed in a global fit for the presence of very light sterile neutrinos, we consider the Neutrino Factory (NF). This instrument provides excellent sensitivities to the standard oscillation parameters, and it can be used for a self-consistent simulation. 
In the proposed experiment, neutrinos are produced from muon decays in straight sections of a storage ring. The design of the Neutrino Factory has been extensively discussed in international studies, such as in \Refs~\cite{Apollonio:2002en,Albright:2004iw,Bandyopadhyay:2007kx}. The International Neutrino Factory and Superbeam Scoping Study~\cite{Bandyopadhyay:2007kx,Abe:2007bi,Berg:2008xx} has laid the foundations for the currently ongoing Design Study for the Neutrino Factory (IDS-NF)~\cite{ids}. This
initiative from about 2007 to 2013 is to present a design report, schedule, cost estimate, and risk assessment 
 for the Neutrino Factory. IDS-NF defines a baseline setup of a high energy neutrino factory with $E_\mu=25 \, \mathrm{GeV}$ and two 
baselines $L_1 \simeq 4 \, 000 \, \mathrm{km}$ and $L_2 \simeq 7 \, 500 \, \mathrm{km}$ 
(the ``magic'' baseline~\cite{Huber:2003ak}) operated by two racetrack-shaped storage rings, where the muon energy is 25~GeV 
(for optimization questions, see, \eg, 
\Refs~\cite{Barger:1999fs,Cervera:2000kp,Burguet-Castell:2001ez,Freund:2001ui,Donini:2005db,Huber:2006wb,Gandhi:2006gu,Kopp:2008ds}).
A key component is the magnetized iron detector (MIND) as far detector, where the magnetization is necessary to distinguish 
the ``right-sign'' (\eg, from $\nu_\mu \rightarrow \nu_\mu$) from the ``wrong-sign'' (\eg, from $\bar\nu_e \rightarrow \bar\nu_\mu$)
muons. The identification of the muon charge of the wrong-sign muons allows for CP violation measurements in the muon neutrino 
appearance channels~\cite{Cervera:2000kp,Burguet-Castell:2001ez}. Physics with near detector configurations for cross section and 
flux measurements have been discussed in \Refs~\cite{Abe:2007bi,Tang:2009na}. Note that recently a smaller scale low energy version 
of a neutrino factory is also been proposed with $E_\mu \sim5$ GeV in 
\Refs~\cite{Geer:2007kn,Bross:2007ts,Huber:2007uj,Bross:2009gk,FernandezMartinez:2010zza,Tang:2009wp}. 
The Neutrino Factory is claimed to be a precision instrument not only because it can 
answer the unknown questions addressed above, but also because it can tell us the story beyond three flavor neutrino oscillation 
physics. Examples are unitarity violation of the mixing matrix coming from heavy fermion 
singlets~\cite{Antusch:2006vwa,FernandezMartinez:2007ms,Abada:2007ux,Goswami:2008mi,Malinsky:2009gw,Meloni:2009cg} and 
non-standard interactions during neutrino productions, propagation in matter, or 
detection~\cite{Huber:2002bi,Kopp:2008ds,Antusch:2008tz,Blennow:2008ym,Ohlsson:2008gx,Malinsky:2008qn,Gavela:2008ra,Meloni:2009cg}. In this study, we focus on the constraint of light sterile neutrinos at the Neutrino Factory, for earlier studies,
see \Refs~\cite{Donini:1999jc,Kalliomaki:1999ii,Donini:2001xp,Donini:2001xy,Dighe:2007uf,Donini:2008wz,Goswami:2008mi,Giunti:2009en,Yasuda:2010rj}. These are either based on the short-baseline (large $\Delta m_{41}^2$ dominance) or long baseline (large $\Delta m_{41}^2$ effects average out) limit. We combine these approaches in one analysis, including possible near detectors and additional channels. Note that in certain short-baseline channels CPT invariance tests are possible~\cite{Giunti:2009en}, as they may be motivated by the recent MiniBooNE results~\cite{AguilarArevalo:2010wv}.

The structure of the paper is organized as follows. In \Sec~\ref{sec:analytic} we illustrate and discuss the parametrization of the neutrino mixing matrix and show approximate expressions for four generation neutrino oscillation probabilities. Matter effects are taken into account in the long baseline setup. We also discuss the experiment simulation. In \Sec~\ref{sec:results}, we show the most general limits on the new parameters without any additional assumptions. A comparison with earlier studies using specific assumptions and a discussion of related issues can be found in \Sec~\ref{sec:comparison}. Finally, we summarize in \Sec~\ref{sec:summary}.

\section{Analytical discussion and simulation techniques}
\label{sec:analytic}

In this section, we first discuss the parametrization of the mixing matrix. Then we derive analytical formulas useful for a first qualitative discussion. Finally, we describe our simulation techniques and assumptions.

\subsection{Mass schemes and parametrization of the mixing matrix}
\label{sec:param}

\begin{figure}[!t]
 \centering
 \includegraphics[height=4cm,width=0.8\textwidth]{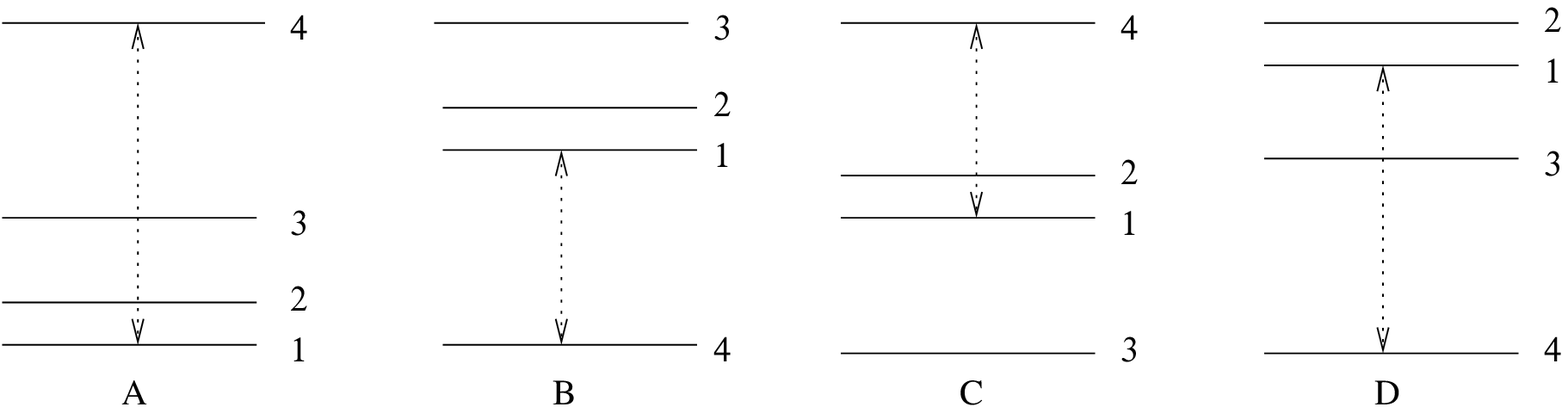}
\mycaption{\label{fig:MH} The mass ordering of four neutrino eigenstates (not to scale). The arrow illustrates the new characteristic mass squared difference $\Delta m_{41}^2$. The four different scenarios correspond to $\ldm>0$, $\vldm>0$ (A), $\ldm>0$, $\vldm<0$ (B), $\ldm<0$, $\vldm>0$ (C), and $\ldm>0$, $\vldm>0$ (D).}
\end{figure}

We study the constraints to the general four neutrino scheme without any assumptions at the Neutrino Factory. While the global fits imply the mixings between active and sterile neutrinos must be small, the four neutrino scheme has to recover the standard picture once we switch off the small mixings between active and sterile neutrinos. The four neutrino schemes can be categorized into two different classes: the 2+2 scheme, in which the  solar and atmospheric mass squared splittings are separated by a new splitting, and the 3+1 scheme, in which the new mass eigenstate is added somewhere to the existing mass pattern. The 2+2 scheme is, at least for an LSND-like new mass squared splitting, strongly disfavored by global fits~\cite{Maltoni:2002ni,Maltoni:2004ei}. The 3+1 scheme, on the other hand, naturally recovers the standard picture in the case of small mixings. Therefore, naturally, we consider the 3+1 scheme only.  We show the possible mass ordering of four neutrino eigenstates in \figu{MH}. The arrow illustrates the new characteristic mass squared difference $\Delta m_{41}^2$. The four different scenarios correspond to $\ldm>0$, $\vldm>0$ (A), $\ldm>0$, $\vldm<0$ (B), $\ldm<0$, $\vldm>0$ (C), and $\ldm>0$, $\vldm>0$ (D). Unless noted explicitly, we show the results for scenario (A).

As far as the parametrization of the four flavor mixing matrix $U_{\alpha i}$ is concerned, the straightforward choice is a parametrization-independent approach. For example, some of the oscillation probabilities in the short-baseline (one mass squared dominance) limit $|\Delta_{41}| \sim \mathcal{O}(1) \gg |\Delta_{31}|$ with $\Delta_{ij} \equiv \Delta m_{ij}^2 L/(4 E)$ read~\cite{Yasuda:2010rj}
\begin{align}
 &\mathcal{P}_{e\mu}= 4 | U_{e4}|^2 |U_{\mu 4}|^2  \sin ^2 \Delta_{41}  \label{equ:pem1} \\
&\mathcal{P}_{e\tau}= 4 | U_{e4}|^2 |U_{\tau 4}|^2 \sin ^2 \Delta_{41} \\
&\mathcal{P}_{\mu \tau}= 4| U_{\mu4}|^2 |U_{\tau 4}|^2 \sin ^2 \Delta_{41} \\
&\mathcal{P}_{\mu\mu}=1- 4 | U_{\mu4}|^2 (1-|U_{\mu 4}|^2) \sin ^2 \Delta_{41}  \label{equ:pmm1}
\end{align}
Information on the combination $ 4| U_{\mu4}|^2 |U_{\tau 4}|^2 $, for instance, can be easily obtained from the NOMAD~\cite{Astier:2001yj} and CHORUS~\cite{Eskut:2007rn} experiments. The mixing matrix elements $U_{\mu4}$ and $U_{\tau 4}$ can then be interpreted in any arbitrary parametrization of $U$. Suppose now we wanted to constrain each mixing matrix element $|U_{\alpha i}|^2$ individually.  It is obvious from the above equations that  $|U_{\mu4}|^2$ could be constrained by $\mathcal{P}_{\mu\mu}$.  The same applies to $|U_{e4}|^2$ by $\mathcal{P}_{ee}$. On the other hand, let's take a look at $|U_{\tau4}|^2$. It would be directly accessible by $\mathcal{P}_{\tau \tau}$, which is, of course, basically inaccessible by any experiment. The alternatives are $\mathcal{P}_{e\tau}$ and $\mathcal{P}_{\mu \tau}$, where $|U_{\tau4}|^2$ enters as a product with $| U_{e4}|^2$ and $| U_{\mu4}|^2$, respectively. If one simulates $|U_{\tau4}|^2=0$ to obtain the expected bound for $|U_{\tau4}|^2$, the marginalization over these two other elements, ending up at zero, will inevitably destroy the sensitivity. Therefore, unless $| U_{e4}|^2>0$ or $| U_{\mu4}|^2>0$ in Nature, one cannot obtain a bound on the individual $|U_{\tau4}|^2$ from these short-baseline probabilities. In summary, one cannot constrain all of the individual elements at the short baseline. We will recover this parametrization-independent statement in our parametrization below.

Another complication of the parametrization-independent approach is that it does not allow for the simultaneous treatment of short and long baselines, and for arbitrary values of $\vldm$. For example, the mixing matrix element combinations in the above equations obviously come from the CP conserving part of the probabilities and rely on the fact that one mass squared splitting is much larger than the other two. If arbitrary $\vldm$ are allowed, other combinations of mixing matrix elements enter, and, in principle, the individual elements are needed separately even in the absence of CP violation. Furthermore, charged and neutral current matter effects at the long baselines complicate this picture even further, see \Ref~\cite{Yasuda:2010rj}. Choosing a particular parametrization for the unitary mixing matrix is therefore the commonly used approach, even in the three flavor picture. Compared to using the mixing matrix elements individually, it has the advantage that unitarity is automatically conserved. Of course, the results are then parametrization-dependent, but the qualitative features are the same as in the parametrization-independent approach, as we will show below.

In principle, there are many different parametrization of the neutrino mixing matrix, since the order of the sub-rotations is arbitrary. For the sake of simplicity, we impose the following requirements:
\bi
 \item
  The standard PMNS matrix has to be recovered in the case of small new mixing angles; that fixes the order of the corresponding sub-sector rotations
 \item
  The phases are attached to the 12-, 13- and 23-rotations. Therefore, if one of the standard mixing angles can be rotated away in a particular measurement, the corresponding phase also automatically becomes unphysical.
 \item
  The order of the 34-24-14-rotations is arbitrary. We choose the 34-angle as the left-most one, which makes it hardest to observe (it affects the $\nu_\tau$-$\nu_s$-mixing). Changing the order here does not change the fact that one of the rotations is difficult to extract.
\ei
Our parametrization explicitly reads
\begin{equation}
    \label{equ:3+1param1}
    U =
    R_{34}(\theta_{34} ,\, 0) \; R_{24}(\theta_{24} ,\, 0) \;
    R_{14}(\theta_{14} ,\, 0) \;
    R_{23}(\theta_{23} ,\, \delta_3) \;
    R_{13}(\theta_{13} ,\, \delta_2) \; 
    R_{12}(\theta_{12} ,\, \delta_1) \,.
\end{equation}
In \equ{3+1param1}, $R_{ij}(\theta_{ij},\ \delta_l)$ are the complex
rotation matrices in the $ij$-plane, defined as:
\begin{equation}
[R_{ij}(\theta_{ij},\ \delta_{l})]_{pq} = \left\{ 
\begin{array}{ll} \cos \theta_{ij} & p=q=i,j \\
1 & p=q \not= i,j \\
\sin \theta_{ij} \ e^{-i\delta_{l}} &	p=i;q=j \\
-\sin \theta_{ij} \ e^{i\delta_{l}} & p=j;q=i \\
0 & \mr{otherwise.}
\end{array} \right.
\end{equation}
This means that $\delta_2$ becomes $\deltacp$ in the three flavor limit.
Note that the order between the 14-and 23-rotations is arbitrary, since these matrices commute.  We do not involve any Majorana phases for their absence in the oscillation probabilities. The parametrization in \equ{3+1param1} is the same as the one used in \Refs~\cite{Donini:2007yf,Donini:2008wz}, and the same as the one in \Ref~\cite{Adamson:2010wi}, since they fix their $\delta_2$, corresponding to our $\delta_1$, to zero.

\subsection{Analytical considerations}
\label{sec:ana}

The general evolution equation of flavor eigenstates in matter can be expressed as
\begin{equation}
 i\frac{d}{dt}|\nu_\alpha\rangle=\mathcal{H}_{\alpha\beta}|\nu_{\beta}\rangle
=\frac{1}{2E}(UD^2U^\dagger+\mathcal{A})|\nu_{\beta}\rangle\,,
\end{equation}
where $D\equiv\mr{Diag}\{m_1,m_2,m_3,m_4\}$ is the diagonal mass matrix and $\mathcal{A}=\mr{Diag}\{a_e,0,0,-a_n\}$ is the matter potential with $a_e=2\sqrt{2}EG_F n_e$ and $a_n=\sqrt{2}EG_F n_n$ (see, \eg, \Ref~\cite{Giunti:2007ry} for details). Here $a_e$ and $a_n$ are the electron and neutron number density in matter, respectively, with $n_e \simeq n_n$ in Earth matter.  Note that the neutral current matter effect may lead to additional matter-driven effects, which are not easy to capture in an analytical treatment. The perturbative solution of this equation depends on the $\Delta_{41}$-regime. We will discuss different regimes separately.

In the one mass squared dominance (short-baseline) limit $|\Delta_{41}| \sim \mathcal{O}(1) \gg |\Delta_{31}|$,  the matter effects can be safely ignored, and the probabilities read:
\begin{align}
 &\mathcal{P}_{e\mu}= \mathcal{P}_{\mu e} = 4 c_{14}^2 s_{14}^2 s_{24}^2 \sin ^2 \Delta_{41} \label{equ:pem2}  \\
&\mathcal{P}_{e\tau}= 4 c_{14}^2 c_{24}^2 s_{14}^2 s_{34}^2 \sin ^2 \Delta_{41} \\
&\mathcal{P}_{es}= 4 c_{14}^2 c_{24}^2 c_{34}^2 s_{14}^2 \sin ^2 \Delta_{41} \\
&\mathcal{P}_{ee}=1-\sin ^2\left(2 \theta _{14}\right)\sin ^2 \Delta_{41}  \label{equ:pee2}\\
&\mathcal{P}_{\mu \tau}= 4 c_{14}^4 c_{24}^2 s_{24}^2 s_{34}^2 \sin ^2 \Delta_{41} \\
&\mathcal{P}_{\mu s} = 4 c_{14}^4 c_{24}^2 c_{34}^2 s_{24}^2\sin ^2 \Delta_{41} \\
&\mathcal{P}_{\mu\mu}=1-c_{14}^2 s_{24}^2  \left[3+2 c_{14}^2 \cos \left(2 \theta _{24}\right)-\cos \left(2 \theta _{14}\right)\right]\sin ^2 \Delta_{41}  \label{equ:pmm2}
\end{align}
Note that the neutral current rate is proportional to $1-\mathcal{P}_{es}$ for electron neutrinos at the source, and $1-\mathcal{P}_{\mu s}$ for muon neutrinos at the source.
These probabilities correspond to the near detector limit at the Neutrino Factory, where the baseline depends on the sensitive $|\vldm|$. For example, at $d \simeq 18 \, \mathrm{km}$, one has optimal sensitivity for $|\vldm| \sim 1 \, \mathrm{eV^2}$, whereas at $d \simeq 1.5 \, \mathrm{km}$, one has optimal sensitivity for $|\vldm| \sim 10 \, \mathrm{eV^2}$ with the 25~GeV Neutrino Factory (\cf, Fig.~2 in \Ref~\cite{Giunti:2009en}). Here $d$ is the distance to the end of the decay straight, which is related to an effective baseline $L_{\mathrm{eff}} = \sqrt{d (d+s)}$ with $s$ the length of the decay straight if the decays are averaged over the straight~\cite{Tang:2009na}.

Let us now compare these equations with the parametrization independent approach in \Sec~\ref{sec:param}, Eqs.~(\ref{equ:pem1}) to~(\ref{equ:pmm1}), assuming that the new mixing angles are small. Obviously, we recover the same qualitative features: $\theta_{24}$ can be measured by $\mathcal{P}_{\mu \mu}$ in the short-baseline limit, and  $\theta_{14}$  by  $\mathcal{P}_{e e}$.   On the other hand, $\theta_{34}$ only shows up in combination with the other small mixing angles. This means that by the same arguments as in \Sec~\ref{sec:param} in the parametrization-independent approach for $|U_{\tau4}|^2$, no separate limit on $\theta_{34}$ can be obtained at the short baselines. In the long-baseline discussion, this parameter is therefore the most interesting one. Note that apart from the above one mass squared dominance ($|\vldm|$) discussion, also the two mass squared dominance ($|\vldm|$ and $|\ldm|$) at somewhat longer baselines ($L \gtrsim 10 \, \mathrm{km}$ at $E_\mu = 25 \, \mathrm{GeV}$) is interesting, because CP violation effects may be observable there~\cite{Donini:2001xy}. However, we focus on near detectors closer to the source in this study. 

For long baselines,  some of the relevant features of the probability transitions can be well understood using simple 
perturbative expansions in appropriate small dimensionless parameters with $\Delta_{31} = \mathcal{O}(1) \ll \Delta_{41}$, \ie, 
if the effects of the large $| \vldm |$ are averaged out. In particular,
up to the second order in 
$
s_{13},  s_{14},  s_{24},  s_{34}, \hat s_{23}=\sin\theta_{23}-\frac{1}{\sqrt{2}}
$,
and considering
$
\Delta_{21}
$
as small as $s_{ij}^2$,
we obtain \footnote{Notice that similar formulas, but using a different perturbative approach, have been derived in 
\Ref~\cite{Donini:2008wz}.}: 
\begin{align}
\mathcal P_{ee} =& 1- 2 s_{14}^2 -4 s_{13}^2\Delta_{31}^2\, 
\frac{ \sin^2 \left(\Delta_{31}-\Delta_e\right)}{(\Delta_{31}-\Delta_e)^2} \, , \label{peelo} \\
\mathcal P_{e\mu} = &P_{e\tau} = 2 s_{13}^2 \Delta_{31}^2\, 
\frac{ \sin^2 \left(\Delta_{31}-\Delta_e \right)}{(\Delta_{31}-\Delta_e)^2} \, , \label{pemulo} \\
\mathcal P_{\mu\mu} =& \cos^2 (\Delta_{31}) (1 - 2 s_{24}^2) + 8 \hat s_{23}^2\sin^2(\Delta_{31}) + c_{12}^2 \Delta_{12}   \sin(2\Delta_{31}) +   \\
&   2 s_{24} s_{34} \cos \delta_3 \Delta_n  \sin(2\Delta_{31}) - \label{pmumulo}\\ 
& 2 s_{13}^2 \Delta_{31}\cos( \Delta_{31})\,\frac{(\Delta_{31}-\Delta_e) \Delta_e \sin (\Delta_{31})-\Delta_{31}\sin\left(\Delta_{31}-\Delta_e \right)\sin (\Delta_e )}{(\Delta_{31}-\Delta_e)^2} \, ,
 \nn \\
\mathcal P_{\mu\tau} =&\sin^2 (\Delta_{31})(1-8 \hat s_{23}^2-s_{24}^2-s_{34}^2)-c_{12}^2 \Delta_{12} \sin (2\Delta_{31})-\nn \\
& s_{24}s_{34}\sin (2\Delta_{31})\left[2 \Delta_n \cos \delta_3-\sin\delta_3\right] - \label{pmutaulo}  \\ 
& s_{13}^2 \Delta_{31}\sin \Delta_{31} \,\frac{\Delta_{31}
\left\{\sin (\Delta_{31}-\Delta_e) +\sin (\Delta_e )\right\}-2 (\Delta_{31}-\Delta_e) 
\Delta_e \cos (\Delta_{31})}{(\Delta_{31}-\Delta_e)^2} \, ,\nn \\
\mathcal P_{\mu s} =& \frac{1}{2}s_{24}^2 (3+\cos (2 \Delta_{31})) + s_{34}^2 \sin^2 (\Delta_{31} )-s_{24} s_{34}
\sin \delta_3\sin (2 \Delta_{31})  \, , \label{pmuslo}
\end{align}
where we used the short-hand notation $\Delta_{e,n} = a_{e,n} L/(4 E)$. 
From these formulas we learn that at the long baselines, $\theta_{14}$ is difficult to measure, because it shows up to leading order only in $\mathcal{P}_{ee}$ as constant term, which we do not consider at the long baseline. On the other hand, $\theta_{24}$ is best accessible by $\mathcal{P}_{\mu \mu}$ with the first term proportional to $\cos^2 (\Delta_{13})$. From the above equations, the leading sensitivity to $\theta_{34}$ can be expected from $\mathcal{P}_{\mu \tau}$ at a long baseline, as claimed by \Ref~\cite{Donini:2008wz}. However, the statistics in this channel is very low, which means that higher order terms in the other channels could still provide the leading sensitivity. Depending on the performance indicators, one of the problems is also the simultaneous presence of $\theta_{24}$ and $\hat s_{23}$ with the same spectral dependence in the first line of $\mathcal P_{\mu\tau}$. Thus, if $\theta_{24}$ is marginalized over, the sensitivity to $\theta_{34}$ disappears. We will test these effects quantitatively later.

Another interesting case, which can be treated analytically, is $|\Delta_{41}| \ll \Delta_{31}$, \ie, $m_1 \simeq m_4$ such that there is no additional $\Delta m^2$, but only 
combinations of the new mixing angles. This case has recently been discussed by the MINOS collaboration~\cite{Adamson:2010wi}.
Interestingly enough, in this limit we can still obtain a rather simple expression for $P_{\mu\mu}$ with exact $\theta_{34}$  
dependence, although with all other mixing angles set to zero and vanishing $\Delta_{21}$ [see App.~\ref{pmm3} for details, 
Eq.~(\ref{first})]. 
From that formula, we can read off that there is, in principle, sensitivity to $\theta_{34}$ via $c_{34}$ in this limit, 
at least without correlations with the other mixing parameters. For the case $|\Delta_{41}| \sim |\Delta_{31}| \gg \Delta_{21}$, 
we have not found any useful analytical expression. We have to rely on numerical results.

\subsection{Simulation techniques}

We have performed the numerical simulation using the  GLoBES software~\cite{Huber:2004ka,Huber:2007ji}. In order to use more than three neutrino flavors, we have defined a user-defined probability engine with the full four flavor probability including appropriate matter effects, assuming $n_n \simeq n_e$. Instead of giving all details, let us just mention some of the complications. First of all, the 
GLoBES code (since version 3) is based on an efficient method to diagonalize Hermitian 3$\times$3 matrices developed by J.~Kopp~\cite{Kopp:2006wp}. Since we had to diagonalize a 4$\times$4 matrix now, we had to return to a GSL-based method, which is less efficient but easy to implement since the interfaces to GSL still exist from prior versions. Second, in order to avoid aliasing effects in the presence of fast oscillations, we have used the filter feature of GLoBES. And third, we have used the built-in standard minimization method GLB\_MIN\_NESTED\_POWELL for the marginalization, which separates the systematics (using the pull method) and oscillation parameter marginalizations. In the presence of so many additional parameters, the new (but more efficient) method GLB\_MIN\_POWELL has rendered insufficient. An alternative may be the Markov chain interface MonteCUBES~\cite{Blennow:2009pk}.

For our experimental setup, we mostly
follow the International Design Study (IDS-NF)~\cite{ids} baseline
setup 1.0, which consists of $2.5 \times10^{20}$ useful muon decays per
polarity, baseline and year with the parent muon energy $E_\mu =25~{\rm GeV}$.
The total running time is assumed to be 10 years. Two magnetized
iron calorimeters (fiducial mass 50 kton) are assumed at $L=4000~{\rm km}$  and $L=7500~{\rm km}$,
respectively. The description of the Neutrino Factory is based on
\Refs~\cite{ids,Huber:2002mx,Huber:2006wb,Ables:1995wq}. The considered oscillation channels are electron to muon neutrino (appearance channels) and muon to muon neutrino (disappearance channels) oscillations, unless noted otherwise.

For the near detectors (we need two of them if the muons circulate in different directions in the storage ring~\cite{Tang:2009na}), we assume a fiducial mass of $32 \, \mathrm{t}$ and a distance of $d=2 \, \mathrm{km}$ from the end of the decay straight, which corresponds to the effective baseline of 2.28~km. This choice is motivated by a compromise between the sensitivity to as small $\Delta m_{41}^2$ as possible, which prefers long baselines, and the longest possible baseline which seems to be plausible. At this distance, the peak sensitivity is achieved at around $|\vldm| \sim 10 \, \mathrm{eV^2}$ (\cf, Fig.~2 in \Ref~\cite{Giunti:2009en}). Note that the near detectors may, depending on the geometry and depth of the storage ring,  also be built close to the surface if the up-going straight is used. We include electron and muon neutrino appearance and disappearance, which means that we, in principle, require charge identification. The muon neutrino channels are assumed to have at least as good capabilities as in the far detectors. The electron neutrino channels are assumed to have an efficiency of 40\% and a charge mis-identification rate at the level of 1\%, following \Refs~\cite{Huber:2006wb,Bross:2009gk}. Note, however, that the charge identification of electrons strongly depends on the detector technology and is, in general, difficult for high energies since the electrons produce quickly electromagnetic showers. From \equ{pee2}, however, we learn that $\mathcal{P}_{ee}$ has a clean signature for $\theta_{14}$. This could be spoilt by $\mathcal{P}_{\mu e}$ in \equ{pem2} at the short baseline in the absence of charge identification if $\theta_{24}$ was sufficiently large. However, this background is
small for small mixings, and $\theta_{24}$ will be constrained by $P_{\mu \mu}$, for which charge identification should be easier to achieve. Therefore, charge identification in the electron neutrino channels turns actually out to be not so important, as we have explicitly checked for the sensitivity limits by adding the electron neutrino and antineutrino event rates. 
As far as the beam geometry is concerned, at this distance the near detectors basically see the same spectrum as the far detectors and the effects of averaging over the decay straight are small at this distance~\cite{Tang:2009na}.

In addition, we test the impact of optional  OPERA-inspired (magnetized) Emulsion
Cloud Chamber (ECC) $\nu_\tau$-detectors as near or far detectors in order
to test the impact of the $\nu_\tau$ channels. The $\nu_{e} \to \nu_\tau$ channel description
is based on \Refs~\cite{Autiero:2003fu,Huber:2006wb}. The $\nu_{\mu}
\to \nu_\tau$ channel is assumed to have the same characteristics as
in \Ref~\cite{Autiero:2003fu}, governed by
\Refs~\cite{Donini:2008wz,FernandezMartinez:2007ms}. Since we assume
that the hadronic decay channels of the $\tau$ can be used as well,
we assume a factor of five higher signal and background than in
\Ref~\cite{Autiero:2003fu}, \ie, 48\% detection efficiency. 
In addition, we assume a detector mass of 10~kt, which is a quite
aggressive choice to see where $\nu_\tau$ detection may be really
relevant.  Note that is yet unclear if an ECC can be operated as close as 2~km
to the Neutrino Factory because of the high scanning load. Therefore,
alternative technologies may be preferable, such as a silicon vertex
detector. In this case, other challenges have to be approached, such 
as the background from anti-neutrino charm production. These issues
are currently under discussion within the IDS-NF.

As far as systematics are concerned, any oscillation signals in the
near detectors, especially in the electron and muon neutrino disappearance channels,
are problematic in the presence of cross section uncertainties.
Therefore, in \Ref~\cite{Giunti:2009en}, a set of near detectors at different
baselines was proposed, similar to the Double CHOOZ or Daya Bay reactor experiments.
With such a combination, cross section errors can be well controlled.
Here we do not simulate such additional detectors explicitly, but only
assume effective normalization errors of 2.5\% (either from external
measurements or from a near detector system).

The best-fit oscillation parameters are taken as follows~\cite{GonzalezGarcia:2010er}:
\begin{align}
&\theta_{12}=34.4^\circ\,,\quad\theta_{13}=5.6^\circ\,,\quad\theta_{23}=42.8^{\circ}\,,\quad
\quad\theta_{i4}=0^{\circ};
\nonumber\\
&\Delta m_{21}^2=7.59\times10^{-5}\text{eV}^2\,,\quad\Delta m_{31}^2=2.46\times10^{-3}\text{eV}^2\,,
\quad\Delta m_{41}^2=1\text{ eV}^2\, .
\end{align}
We impose external $1\sigma$ errors on $\Delta m^2_{21}$ (4\%) and
$\theta_{12}$ (4\%) and on $\Delta m^2_{31}$ (10\%) and
$\theta_{23}$ (10\%) as conservative estimates for the current
measurement errors~\cite{GonzalezGarcia:2010er}. 
We also include a 5\% matter density uncertainty~\cite{Geller:2001ix,Ohlsson:2003ip}.

\section{Generalized exclusion limits}
\label{sec:results}

In this section, we discuss general constraints to the new mixing angles $\theta_{14}$, $\theta_{24}$, and $\theta_{34}$, and the additional mass squared difference $\vldm$ {\em without any additional assumptions}. In particular, we do not assume that $\vldm$ is in a particular range, such as the LSND-motivated one which leads to averaging at the long baselines. In addition, we do not assume that some of the not shown parameters take particular fixed values.

As performance indicator, we use the sensitivity to $\theta_{ij}$ similar to the CHOOZ limit for the $\theta_{13}$-$\ldm$ plane. We compute the simulated rates with $\theta_{14}=\theta_{24}=\theta_{34}=0$ and $\vldm=0$, corresponding to the hypothesis of no effect of an additional sterile neutrino. Then we can discuss several exclusion limits for the new mixing angles, with the unknown parameters marginalized over. The exclusion limit for each new mixing angle will, in general, depend on $\vldm$ similar to the CHOOZ limit. Therefore, we show three different exclusion planes $\theta_{i4}$-$\vldm$. Note that the sensitivity for any combination $\theta_{i4}$-$\theta_{k4}$ ($i \neq k$) will typically vanish, since $\vldm$ is marginalized over. For a comparison to the existing literature, see the next section.

\begin{figure}[!t]
 \centering
\includegraphics[width=0.32\textwidth]{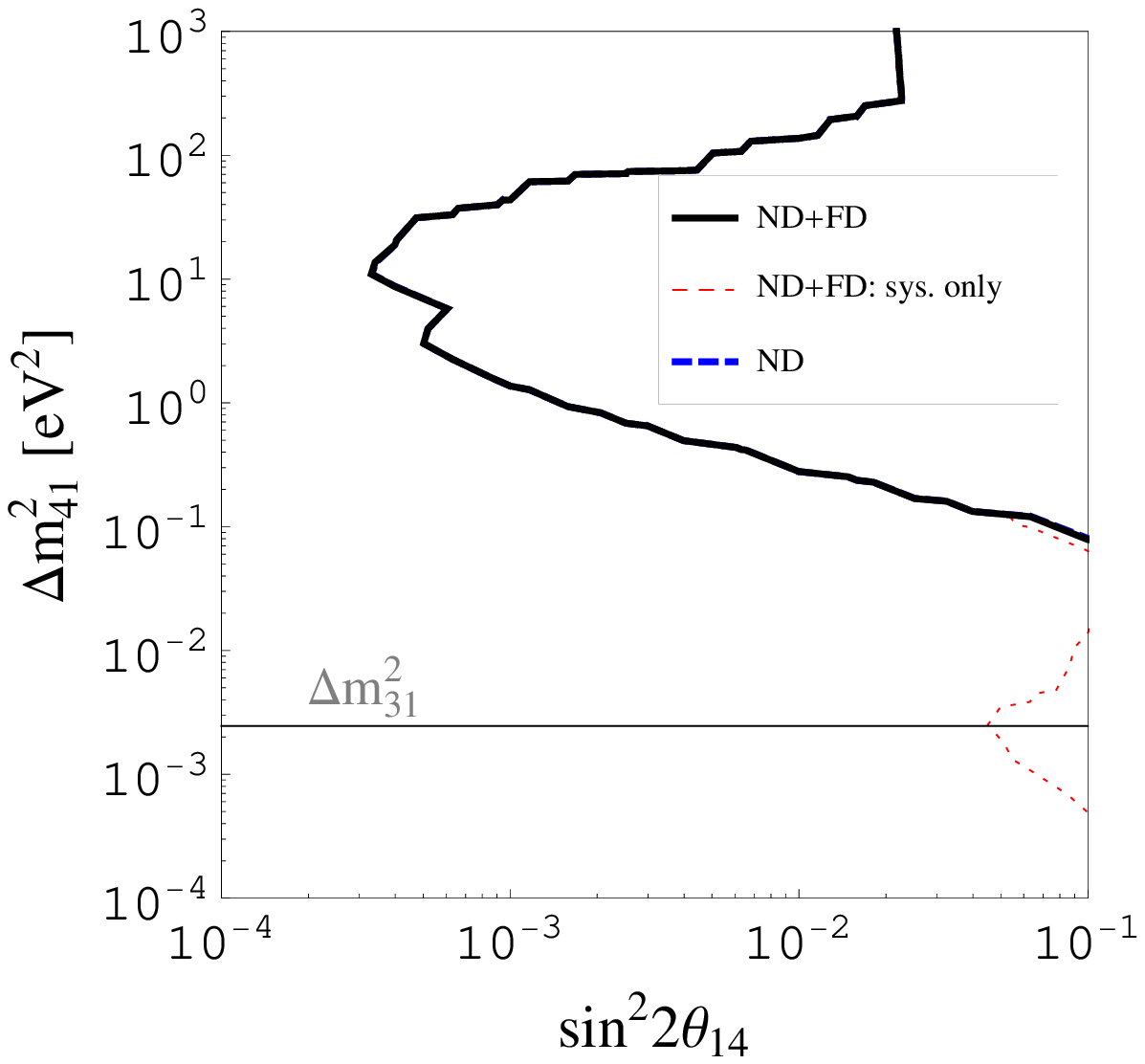} \includegraphics[width=0.32\textwidth]{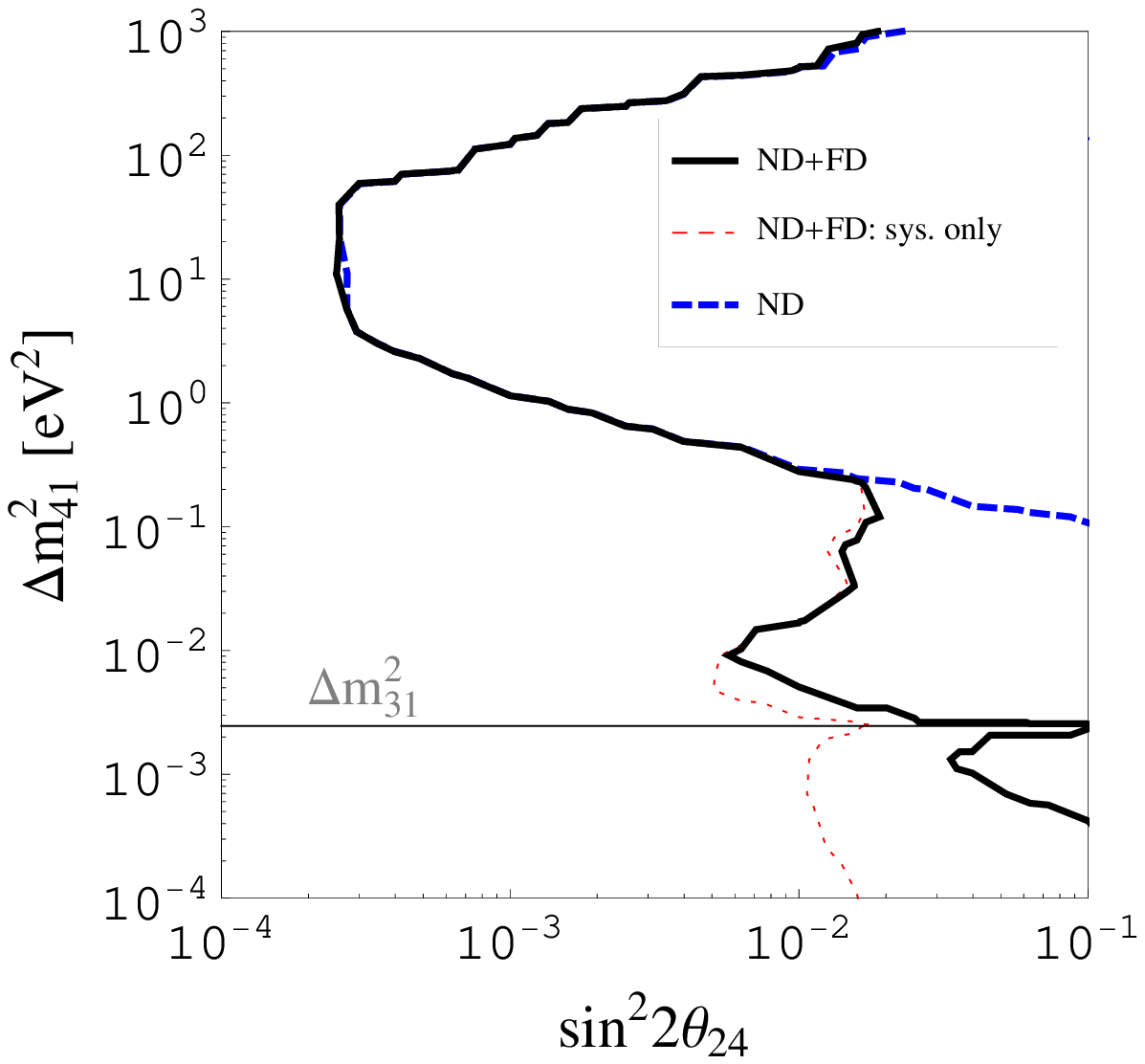} \includegraphics[width=0.32\textwidth]{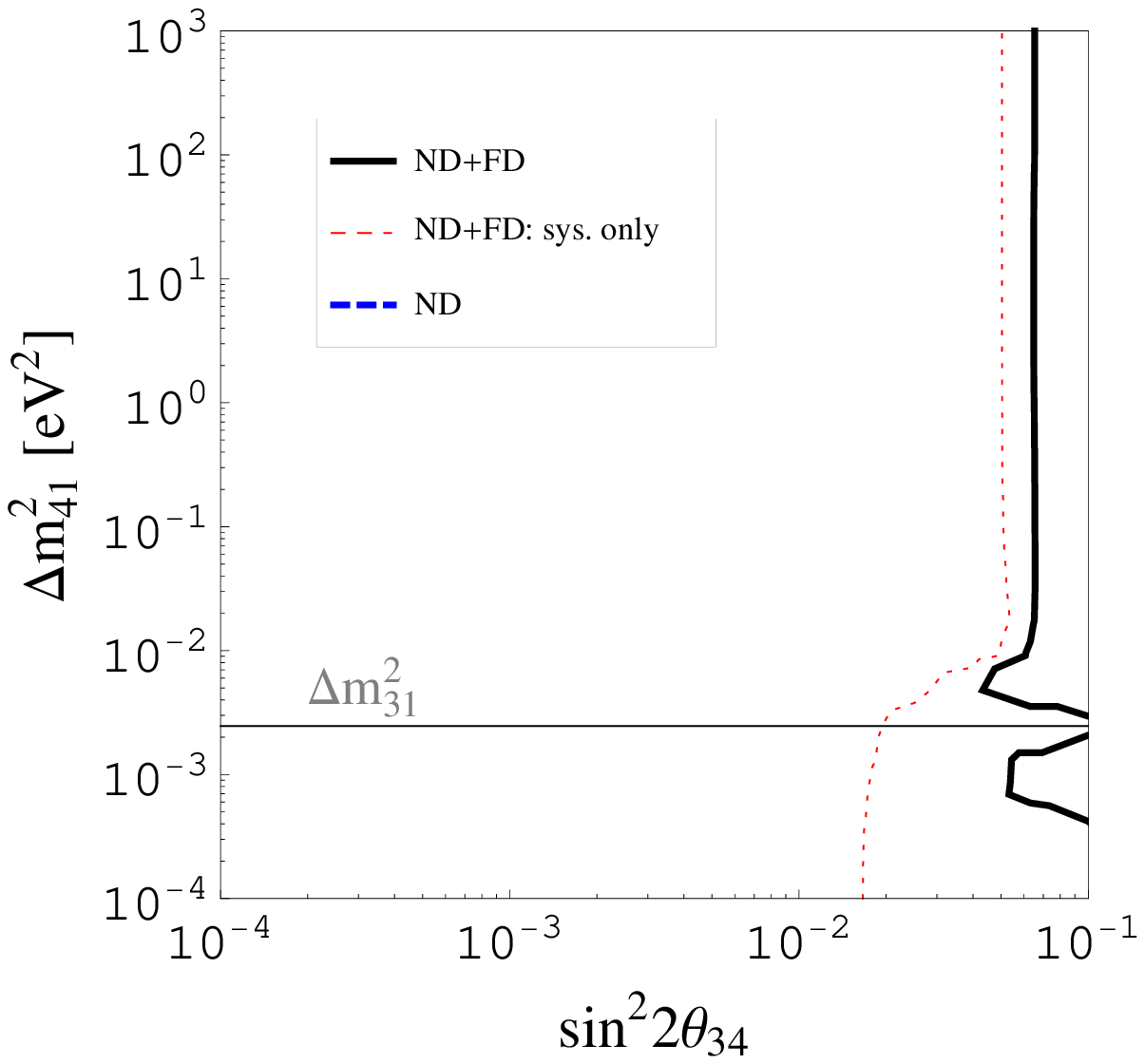}
\mycaption{\label{fig:theta-mass}The exclusion limit for $\sin^22\theta_{i4}$--$\Delta m_{41}^2 (i=1,2,3)$ (region on r.h.s. of curves excluded). Here gives results of the standard IDS-NF ($4000$ km and $7500$ km) setup with near detectors (thick solid curves), and the impact of the near detectors is also shown separately (thick dashed curves). The thin dashed curves only consider systematics. Mass ordering~A assumed (\cf, \figu{MH}), 90\% CL (2 d.o.f).}
\end{figure}

Our main result can be found in \figu{theta-mass}: The exclusion limit for $\sin^22\theta_{i4}$--$\Delta m_{41}^2 (i=1,2,3)$ (region on r.h.s. of curves excluded). Let us first of all discuss the effect of the near detectors separately (thick dashed curves). Obviously, the main sensitivity is obtained at about $\Delta m_{41}^2 \simeq 10 \, \mathrm{eV}^2$, which comes from the distance chosen for the near detectors. By changing the near detector locations, the position of the main peak can be controlled. If required by the recent MiniBooNE results, a longer baseline than 2~km may have to be chosen. As discussed in \Sec~\ref{sec:ana}, the main sensitivity to $\theta_{14}$ comes from electron neutrino (antineutrino) disappearance and the main sensitivity to $\theta_{24}$ from muon neutrino (antineutrino) disappearance. Since the efficiencies for muon neutrino detection are typically better, the sensitivity to $\theta_{24}$ is slightly better than that to $\theta_{14}$ for our assumptions. As expected, there is no sensitivity to $\theta_{34}$ coming from the near detectors, because the $\nu_\tau$ disappearance channel does not exist. 
Note that the appearance channels at short baselines always depend on the product of two new parameters, \cf, \Sec~\ref{sec:ana}. This means that no sensitivity can be obtained from the appearance channels without additional assumptions. Consider, for instance, $\mathcal P_{e \mu}$ in \equ{pem2}, which is sensitive to the product of $\theta_{14}$ and $\theta_{24}$. In order to constrain $\sin^22\theta_{14}$--$\Delta m_{41}^2$, $\theta_{24}$ is to be marginalized over, which destroys the sensitivity. The problem is that this channel only limits a combination of $\theta_{14}$, $\theta_{24}$, and $\Delta m_{41}^2$, which is difficult to display.  Therefore, by our definition, we do not find any sensitivity from this channel, as opposed to Fig.~7 in \Ref~\cite{Donini:2001xy}.

For the effect at the long baselines, it is first of all useful to consider the thin dashed curves with systematics only. In all three panels, the sensitivity changes as a function of $\vldm$ in the region where $\vldm \sim \ldm$. It comes from the fact that the Neutrino Factory is sensitive to the atmospheric oscillation frequency, whereas for $\vldm \sim \sdm$, no particular additional effects from the solar frequency can be found. As expected [\cf, Eq.~(\ref{pmumulo})], the main sensitivity is found for $\theta_{24}$ which can be measured with the $\mathcal P_{\mu \mu}$ disappearance channel. However, there is also some sensitivity to $\theta_{14}$, which vanishes after the marginalization, and some sensitivity to $\theta_{34}$, which is even present for $\vldm=0$ for systematics only (\cf,  \App~\ref{pmm3} for analytical formulas in that limit). After marginalization (thick solid curves), only the sensitivities to $\theta_{24}$ and $\theta_{34}$ remain in the $\vldm$ regions close to the atmospheric $\ldm$ and above, where the effects of $\vldm$ average out. Very interestingly, note that mixing angle correlations destroy the sensitivities for $\vldm=\ldm$, where $m_4=m_3$ and no additional $\vldm$ is observable, leading to small gaps (see horizontal lines). The sensitivity to $\theta_{34}$ is not visible in \Sec~\ref{sec:ana} even for large $\vldm$. We have tested that it is a matter potential-driven, statistic limited higher order effect in $\theta_{34}$, present in the muon neutrino disappearance channels.  Note that while one may expect some effect from $\mathcal P_{\mu \tau}$, we have tested the impact of $\nu_\tau$ detectors at all baselines, and we have not found any improvement of the sensitivities. The reason is that the $\theta_{34}$ effect at the long baseline comes with the same energy dependence as the $\theta_{24}$ effect, which means that one cannot disentangle these, as discussed in \Sec~\ref{sec:ana}.

\begin{table}[t]
\begin{center}
\begin{tabular}{r|rrr|rrr}
\hline
$\vldm$ [eV$^2$] & $\sin^2 2 \theta_{14}$ & $\sin^2 2 \theta_{24}$ & $\sin^2 2 \theta_{34}$ 
& $\theta_{14} [^\circ]$& $\theta_{24}[^\circ]$ & $\theta_{34}[^\circ]$ \\
\hline
$0.001$ & $0.403$  & $0.029$  & $0.042$ & $19.7$& $4.9$& $5.9$\\
$0.01$  & $0.224$  & $0.004$  & $0.044$ & $14.1$& $1.9$& $6.1$\\
$0.1$   & $0.054$  & $0.013$  & $0.047$ & $6.7$ & $3.3$& $6.3$\\
$1$     & $0.001$  & $0.0009$ & $0.047$ & $0.9$ & $0.8$& $6.3$\\
$10$    & $0.0002$ & $0.0002$& $0.047$ & $0.5$ & $0.5$& $6.3$\\
$100$   & $0.0043$ & $0.0006$ & $0.047$ & $1.9$ & $0.7$& $6.3$\\
$1000$  & $0.0168$ & $0.015$  & $0.047$ & $3.7$ & $3.5$& $6.3$\\
\hline
\end{tabular}
\end{center}
\mycaption{\label{tab:limits} Exclusion limits (90\% CL, 1 d.o.f.) for several selected (fixed) values of $\vldm$. Mass ordering~A assumed. The simulation includes both near and far detectors.}
\end{table}

In view of the three panels, it is not easy to disentangle the parameters for arbitrary massive sterile neutrinos. Parameter correlations lead to a pollution of the exclusion limit of a particular mixing angle with $\Delta m_{41}^2$. In addition, there is a competition between $\Delta m_{41}^2$ and $\Delta m_{31}^2$ at the long baseline. Near detectors, on the other hand, have very good sensitivities to $\theta_{14}$ and $\theta_{24}$ but cannot measure $\theta_{34}$. Nevertheless, the absolute values of the sensitivities are quite impressive, see \Tab~\ref{tab:limits}. Especially, $\theta_{24}$ can be very well constrained close to the atmospheric mass squared difference range. This indicates that sterile neutrino bounds in that range should be also obtainable from current atmospheric neutrino oscillation experiments. In addition, solar and supernova neutrino experiments may test even smaller $\vldm \ll 10^{-3} \, \mathrm{eV}^2$, which is to be investigated.

\begin{figure}[!t]
 \centering
 \includegraphics[width=0.32\textwidth]{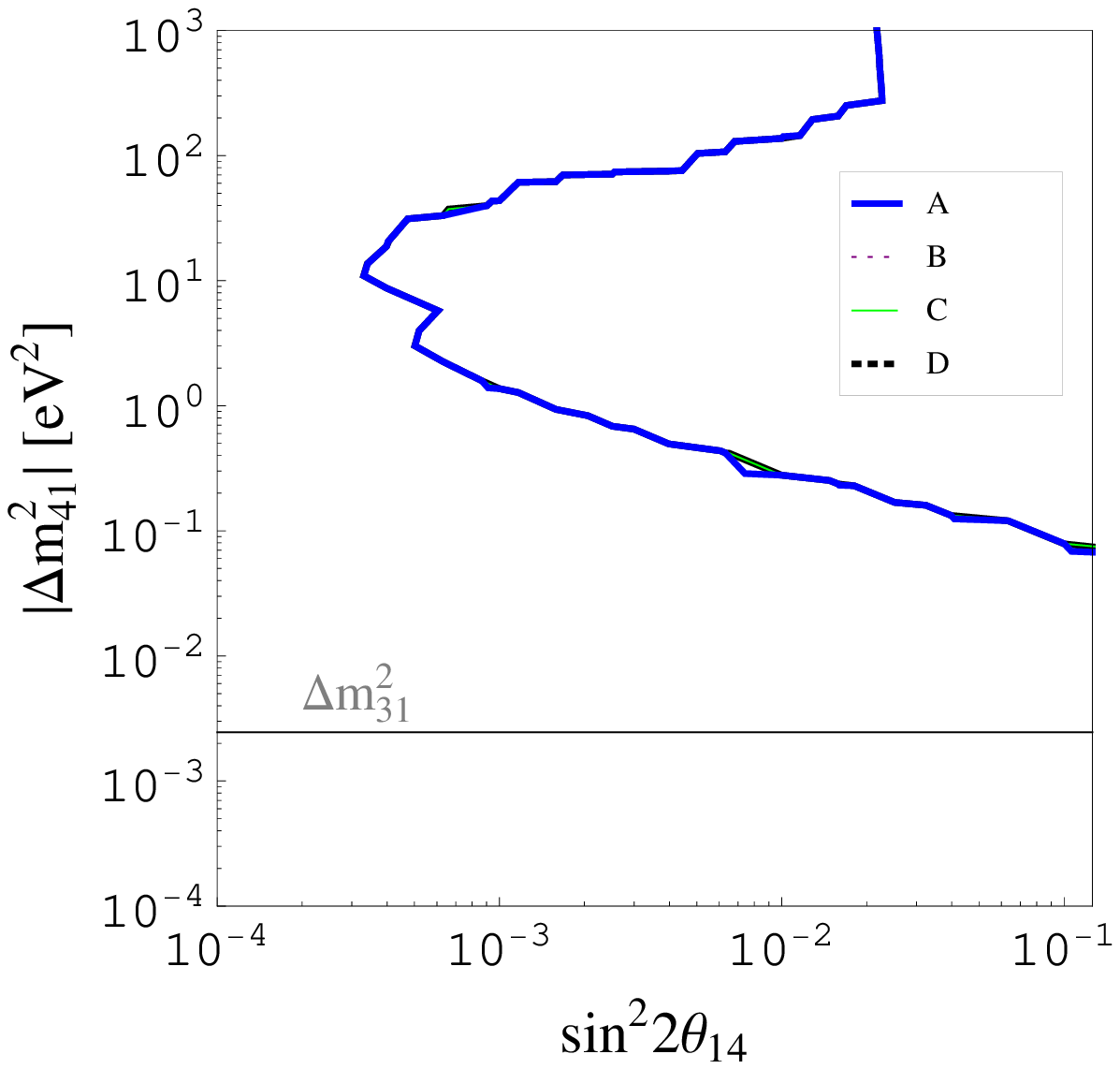} \includegraphics[width=0.32\textwidth]{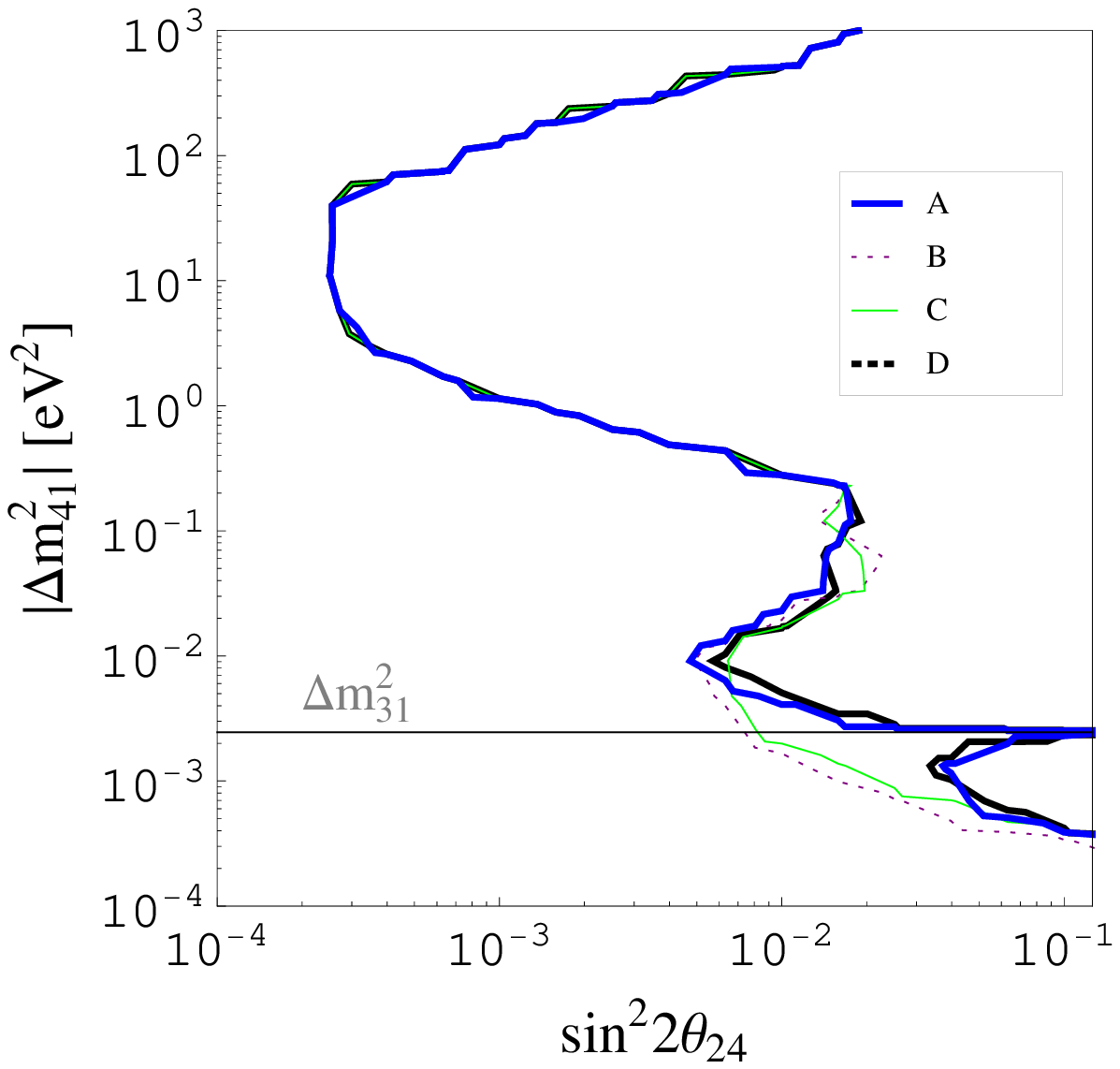} \includegraphics[width=0.32\textwidth]{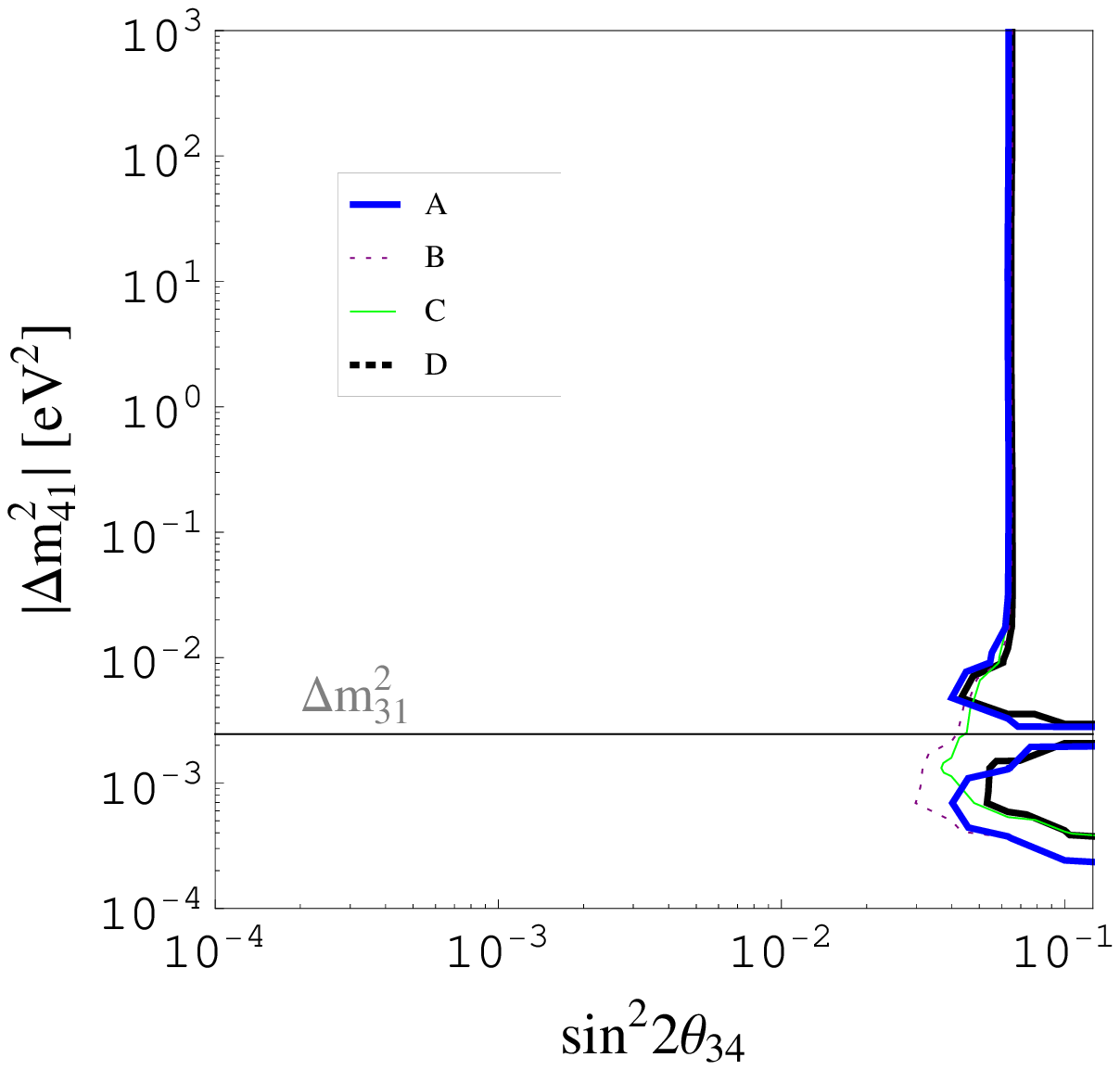}
\mycaption{\label{fig:MH-theta-mass} The exclusion limit for $\sin^22\theta_{i4}$--$\Delta m_{41}^2 (i=1,2,3)$ (region on r.h.s. of curves excluded). Here gives results of the standard IDS-NF ($4000$ km and $7500$ km) setup with detectors for the four different mass orderings in \figu{MH}; 90\% CL (2 d.o.f.).}
\end{figure}

To illustrate the effect from different mass orderings, as shown in Fig.~\ref{fig:MH}, we display these figures for the standard IDS-NF ($4000$ km and $7500$ km) setup in combination with near detectors in Fig.~\ref{fig:MH-theta-mass}. Note that only the absolute value of the new mass squared difference is shown at the vertical axes. The upper peak hardly depends on the mass ordering, as it is obvious from the analytical short-baseline formulas in \Sec~\ref{sec:ana}.  The lower (long-baseline) peak, which is only present in the middle and right panels, somewhat depends on the mass ordering. We identify two qualitatively different cases: In schemes A and D, the sensitivity is destroyed just at the value of $\ldm$. In these cases, \cf, \figu{MH}, mass eigenstates~3 and~4 are on top of each other, which means that there is no additional mass squared difference. The parameter correlations (marginalization over the unknown parameters) then destroy the sensitivity because the new mixing angles cannot be disentangled, in spite of the additional neutral current matter effect. This is different for schemes B and C, for which mass eigenstate~4 is on the opposite site of mass eigenstate~3. Although the absolute values of $\vldm$ and $\ldm$ are similar, these mass squared differences have different signs leading to different (charged current) matter effects.  

\section{Exclusion limits with special assumptions}
\label{sec:comparison}

Here we compare our analysis with different approaches in the literature under special assumptions. The most common assumption for sterile neutrino bounds at the long baselines is $|\Delta m^2_{41}| \sim\mathcal{O}(1)$ eV$^2$, as motivated by LSND, which leads to averaging over the fast $\ldm$ at the long baselines. Another assumption, which we have found in \Ref~\cite{Adamson:2010wi}, is $\vldm \rightarrow 0$. The third case we consider is the two-flavor short-baseline limit. These examples are particularly useful to discuss some subtleties when short- and long-baseline results are to be combined. Note that, for the comparison to the existing literature, we use studies which use the same parametrization of the four neutrino mixing matrix as ours for the sake of simplicity.

\subsubsection*{LSND-motivated $\boldsymbol{\vldm}$}

The first assumption, which we test, is an LSND-motivated $\vldm$. In general, we assume that $|\vldm| \gg |\ldm|$, which leads to the averaging of the fast oscillations at the long baselines. This limit is frequently used for the discussion of long baselines only, \ie, without near detectors, since oscillation effects may be present in the near detectors.

\begin{figure}[!t]
 \centering
 \includegraphics[width=0.5\textwidth]{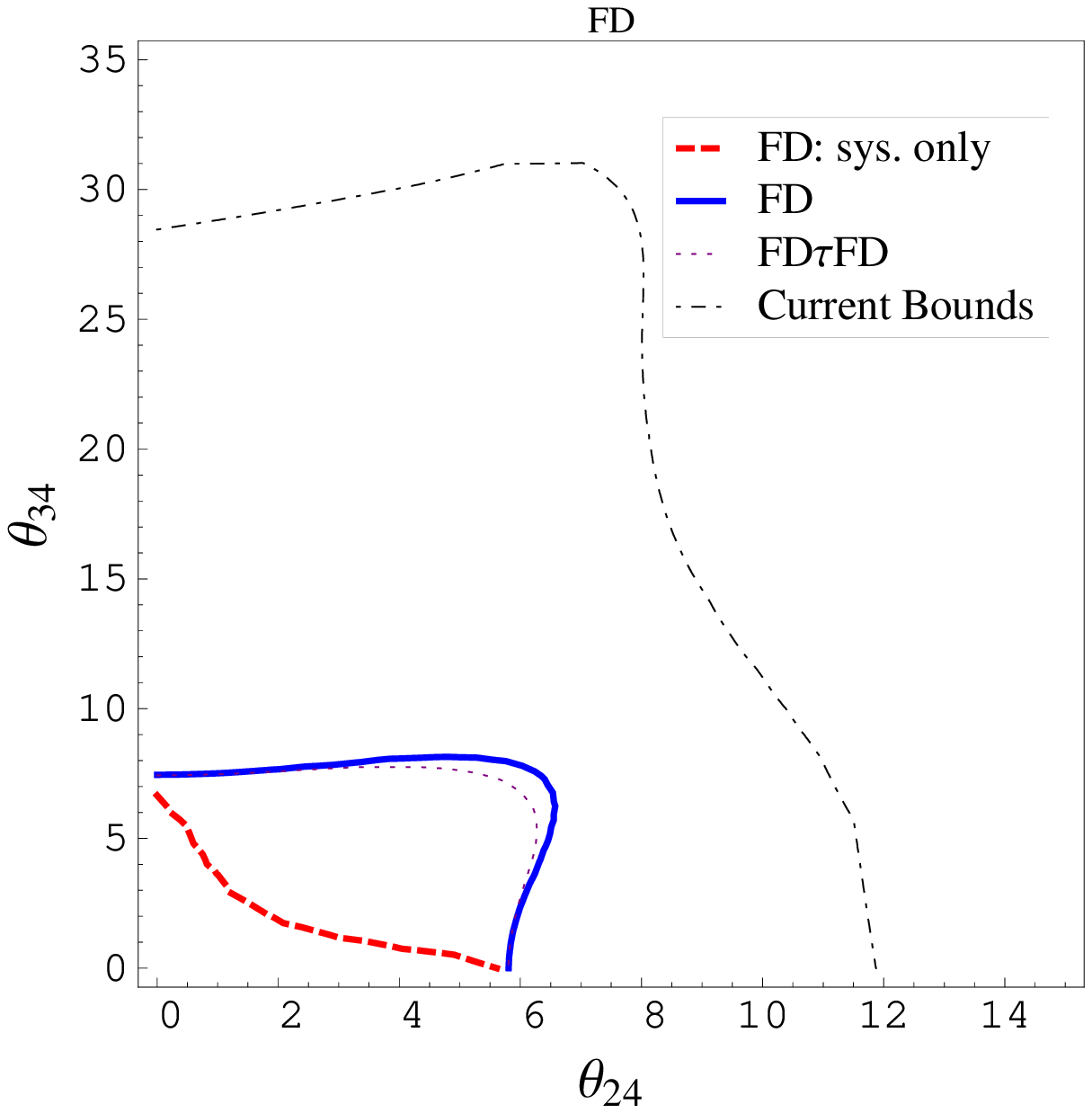}
\mycaption{\label{fig:th24-th34}
The exclusion limit for $\theta_{24}$--$\theta_{34}$ in degrees  (90\% CL, 2 d.o.f.), where $|\vldm| \gg |\ldm|$. We show the result for the IDS-NF setup ($4000$ km and $7500$ km) using far detectors only (thick solid curve). The thick dashed curve corresponds to the result without correlations (systematics only), the dotted curve the result including a $\nu_\tau$ detector at $4000$ km. In addition, the current bounds are shown (from \Ref~\cite{Donini:2007yf}). 
}
\end{figure}

We show in \figu{th24-th34} an example for such an analysis using the far detectors only (thick solid curve), which is to be compared with Fig.~9 in \Ref~\cite{Donini:2008wz} for slightly different parameter values (here the simulated $\theta_{14}=0$). For comparison, the current bound is shown (\cf, Fig.~2 in \Ref~\cite{Donini:2007yf}). One can easily see that the Neutrino Factory could improve the current bounds on $\theta_{24}$ and $\theta_{34}$ by a factor of a few if $|\vldm|$ is assumed to be large. Note, however, that the marginalization over $\vldm$ would lead to vanishing sensitivity.  If the (thick dashed) curve with only systematics is considered, the cross terms in the long-baseline probabilities in \Sec~\ref{sec:ana}, such as in $\mathcal P_{\mu \mu}$ in Eq.~(\ref{pmumulo}),  are switched off, and result improves along the diagonal.  Such cross terms are also present in $\mathcal P_{\mu \tau}$ in Eq.~(\ref{pmutaulo}), sometimes called the ``discovery channel''~\cite{Donini:2008wz}, which leads to a slight improvement if an additional $\nu_\tau$ detector at the intermediate baseline is used (thin dotted curve). However, as we noted above, the main effect on $\theta_{34}$ comes from the disappearance channels, especially of the very long ($7500$~km) baseline. 

\begin{figure}[!t]
 \centering
\includegraphics[width=0.5\textwidth]{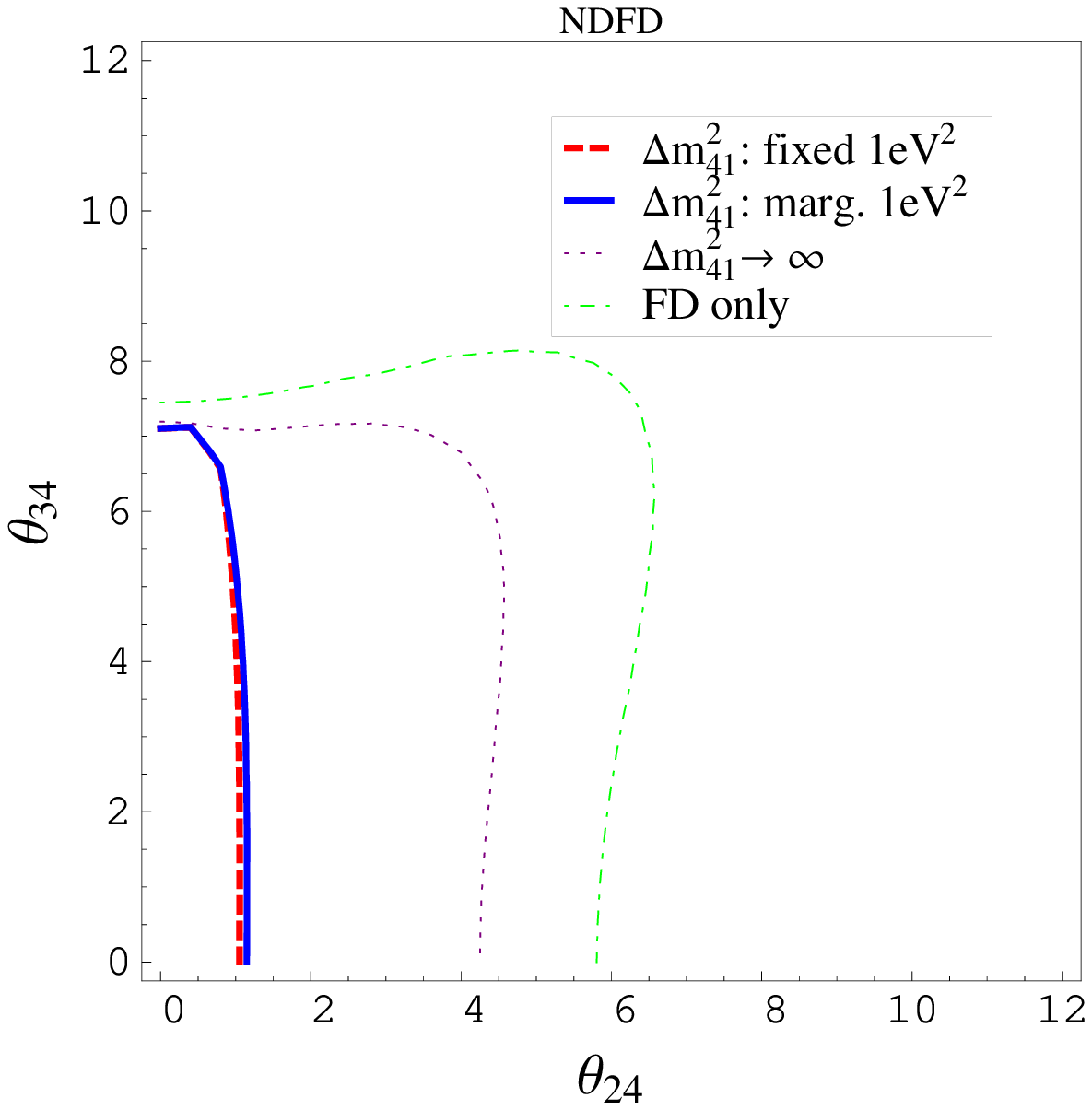}
\mycaption{\label{fig:th24-th34-near}
The exclusion limit for $\theta_{24}$--$\theta_{34}$ in degrees (90\% CL, 2 d.o.f.), where $|\vldm| \gg |\ldm|$. Compared to \figu{th24-th34}, additional near detectors at 2~km are used. This leads to the dependence on the assumptions for $\vldm$ (and $\sin^2 2 \theta_{14}$), as given in the plot legend (see main text for details). Here we assume $\sin^22\theta_{14}=0.01$ for the curves including near detectors. Note that  the dashed-dotted curve corresponds to the thick solid curve in \figu{th24-th34}.  
}
\end{figure}

Let us now test the impact of additional near detectors on this scenario, see \figu{th24-th34-near}. Obviously, we have to be careful, because a $\vldm \sim 1 \, \mathrm{eV}^2$ might cause observable effects in the near detectors, and, in principle, we also have to marginalize over $\vldm$. We especially expect some impact on the measurement of $\theta_{24}$ by the muon neutrino disappearance channel in \equ{pmm2}. If we fix $\vldm = 1 \, \mathrm{eV}^2$, we find considerably better sensitivity including the near detectors, compare the dashed-dotted curve (without near detectors) with the thick dashed curves. However, note that the marginalization over $\vldm$ will in this case destroy the sensitivity again.  This problem can be circumvented by additional assumptions. Consider, for instance, a non-zero value of $\sin^2 2 \theta_{14}$ chosen by Nature. Then $\vldm$ can be actually measured by $\mathcal P_{ee}$ in the near detectors, see \equ{pee2}, and the marginalization over $\vldm$ can be performed. In this case, the impact of the $\vldm$ marginalization is small, as it can be seen from the comparison between the thick dashed and thick solid curves. 
Another plausible assumption may be that $\vldm$ is so large that the effect even averages out in the near detectors. Therefore, we show the thin solid curve for a KeV sterile neutrino with $\vldm = 10^6 \, \mathrm{eV}^2$, which is sufficiently large. Here the near detectors somewhat improve the sensitivity compared to the no near detector case, but the effect is not as large as for the $\vldm \sim 1 \, \mathrm{eV}^2$, $\sin^2 2 \theta_{14}=0.01$ case. The purpose of this example is to illustrate that any combined fit of two new parameters not including $\vldm$, such as the one in \figu{th24-th34}, face subtleties if near and far detectors are combined. Typically, additional assumptions are needed, and the interpretation of the results becomes assumption-dependent. This is in contrast to the figures in the previous section, which do not depend on assumptions.% 

\begin{figure}[!t]
 \centering
 \includegraphics[width=12cm,height=12cm]{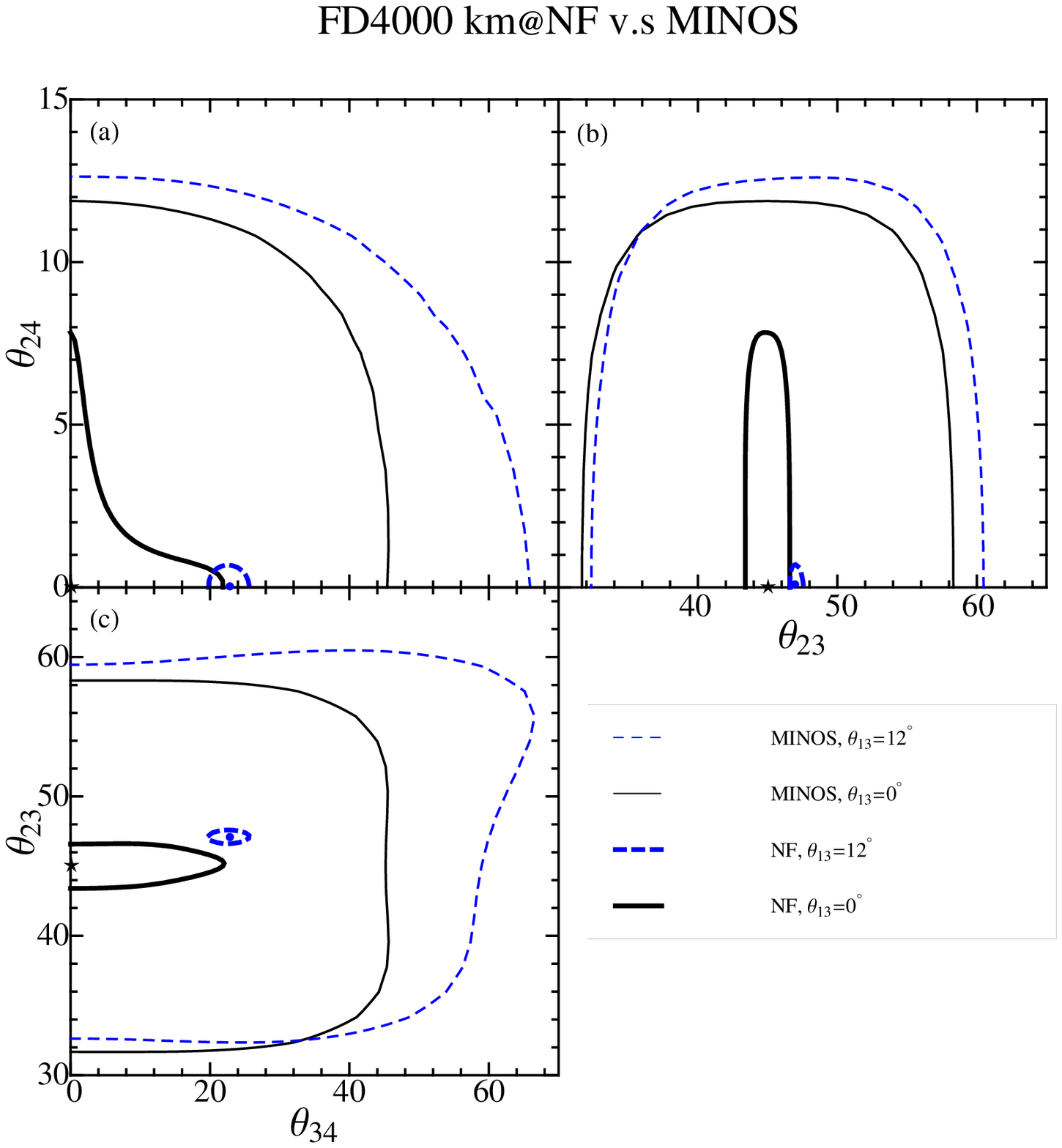}
 % mod1.eps: 0x0 pixel, 300dpi, 0.00x0.00 cm, bb=0 0 300 310
\mycaption{\label{fig:minos-mod2FD} Exclusion limits in the $\theta_{34}$--$\theta_{24}$ (a), $\theta_{23}$--$\theta_{24}$ (b), and $\theta_{34}$--$\theta_{23}$ (c) planes at the 90\% CL. 
The different curves correspond to MINOS and the Neutrino Factory (NF) with $\Delta m_{41}^2=1$ eV$^2$. Here only one NF far detector at 4000~km is used (without near detectors). All contours represent $90\%$ confidence level. The solid curves assume $\theta_{13}=0^\circ$, while the dashed curves assume $\theta_{13}=12^\circ$. The best-fit values are marked in the figure, the one of $\theta_{24}$ is zero. The MINOS curves are taken from \Ref~\cite{Adamson:2010wi}.}
\end{figure}

\begin{figure}[!t]
 \centering
 \includegraphics[width=12cm,height=12cm]{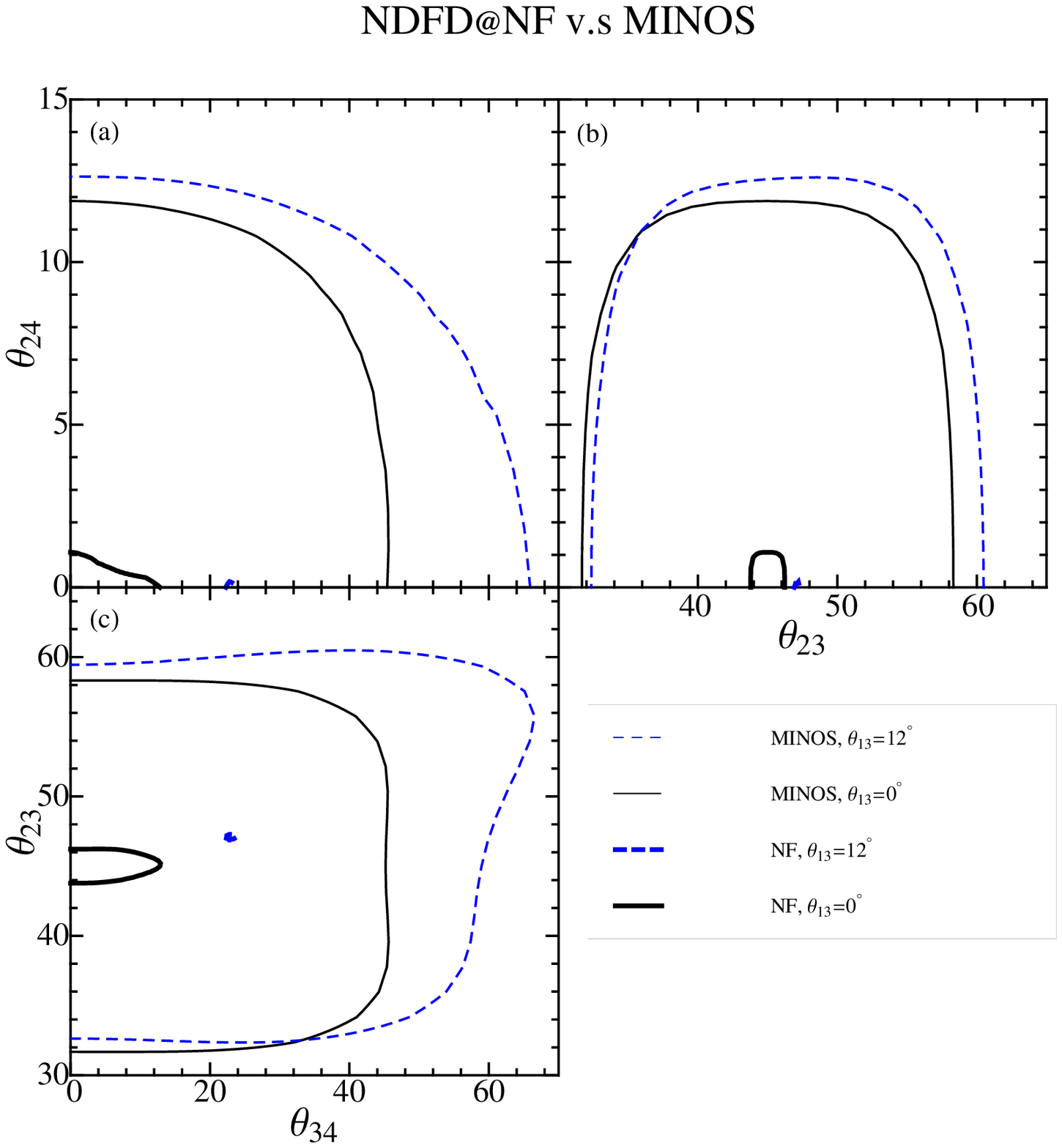}
 % mod1.eps: 0x0 pixel, 300dpi, 0.00x0.00 cm, bb=0 0 300 310
\mycaption{\label{fig:minos-mod2} Same as \figu{minos-mod2FD}, but additional far detector at 7500~km and near detectors at 2~km included. The NF contours for large $\theta_{13}=12^\circ$ almost shrink to points.}
\end{figure}

A similar analysis is performed in \Ref~\cite{Adamson:2010wi} by the MINOS collaboration.  The mixing matrix parametrization in this reference is equivalent to ours since they fix their $\delta_2$, corresponding to our $\delta_1$, to zero. One of the schemes tested in \Ref~\cite{Adamson:2010wi}  assumed that $|\vldm| \gg |\ldm|$ (in mass ordering~A). We show their result for the
combined fits in the $\theta_{34}$--$\theta_{24}$ (a), $\theta_{23}$--$\theta_{24}$ (b), and $\theta_{34}$--$\theta_{23}$ (c) planes in \figu{minos-mod2FD}. Note that the not shown parameters are fixed, such as $\theta_{14}=0$ and the phases, and that $\theta_{13}$ is fixed to two different values (not marginalized over). The best-fit values are also marked.  We also show the results for the Neutrino Factory under the same assumptions, where we use the 4000~km baseline only. Obviously, the Neutrino Factory would reduce the allowed parameter space significantly, especially if $\theta_{13}$ is large. Again, note that some of the parameters are fixed here, and the full marginalization  would destroy the sensitivities.  The impact of additional near detectors, which are especially sensitive to $\theta_{24}$, and an additional far detector at 7500~km, which is sensitive to $\theta_{34}$, is shown in \figu{minos-mod2}. Here the NF contours for large $\theta_{13}=12^\circ$ almost shrink to points, and are hardly visible anymore. In this figure, $\vldm= 1\, \mathrm{eV}^2$ is assumed, and $\vldm$ is not marginalized over. For such a value of  $\vldm$, we would also expect a small effect in the MINOS near detector, which is, however, not considered in \Ref~\cite{Adamson:2010wi}.

\subsubsection*{The special case $\boldsymbol{\vldm \rightarrow 0}$}

\begin{figure}[!t]
 \centering
 $\begin{array}{lll}
 \includegraphics[width=0.48\textwidth]{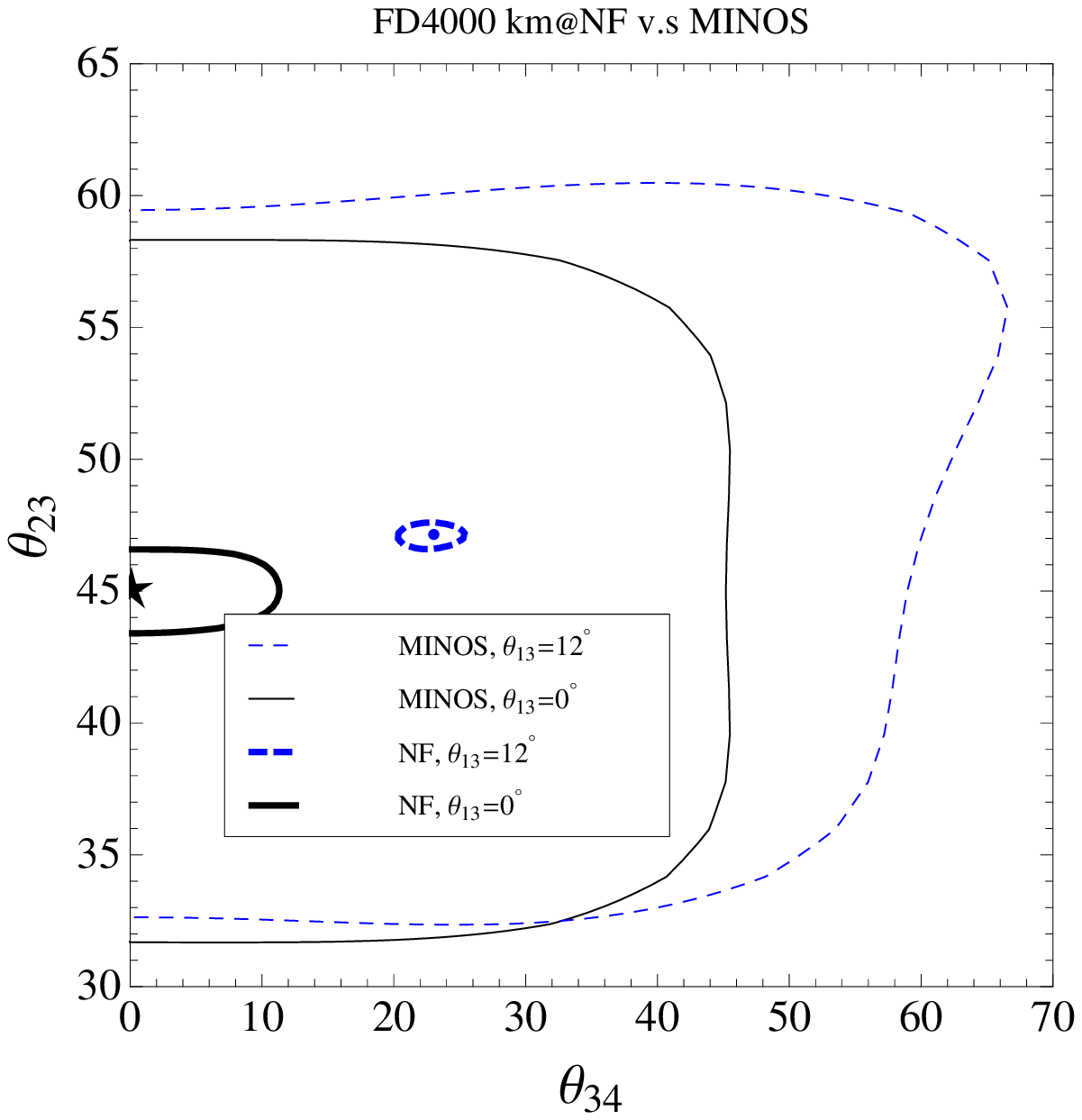}
&\includegraphics[width=0.48\textwidth]{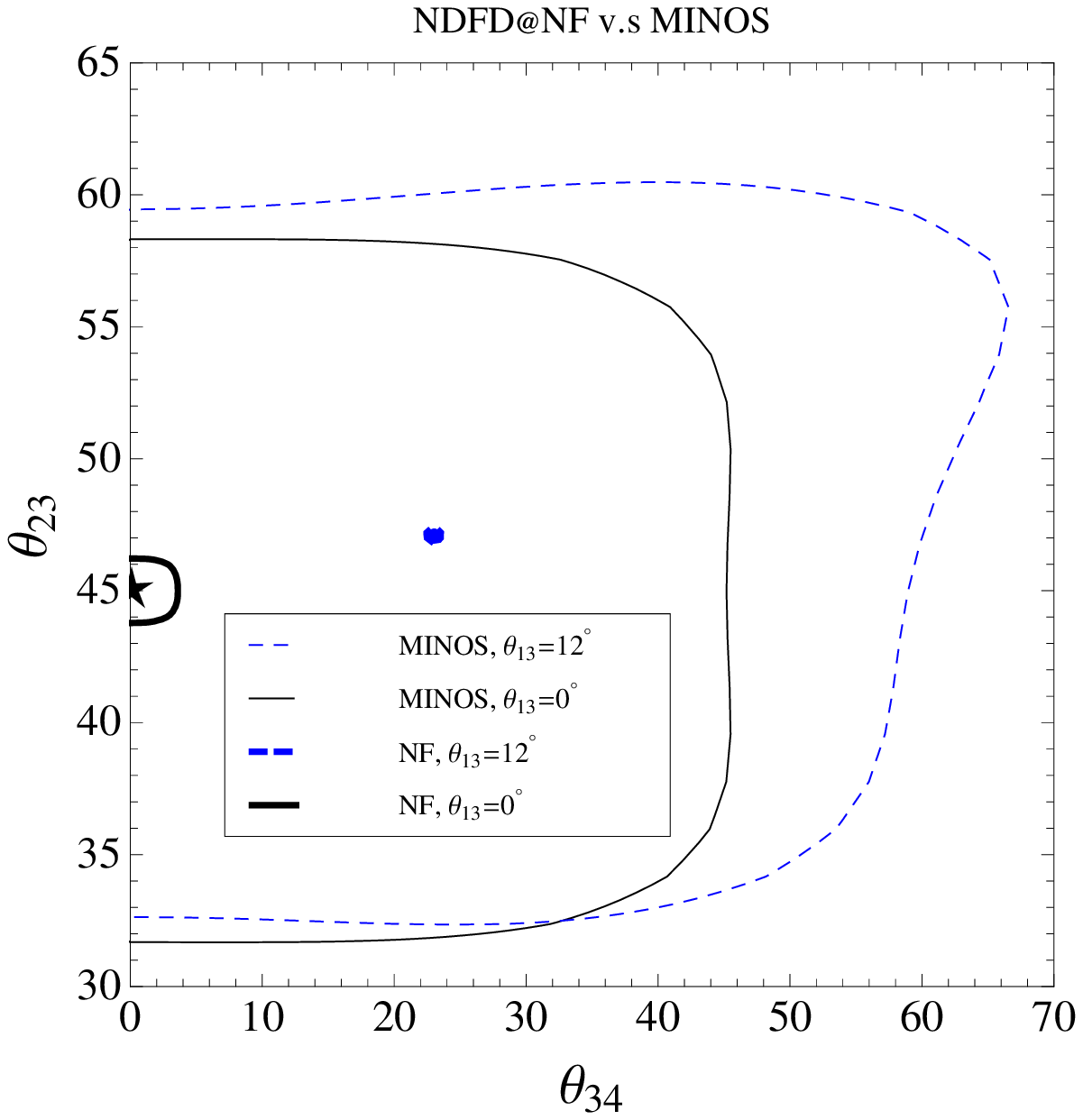}
 \end{array}$
\mycaption{\label{fig:minos-mod1}Exclusion limits in the $\theta_{34}$--$\theta_{23}$ plane at the 90\% CL. 
The different curves correspond to MINOS and the Neutrino Factory (NF) with $\Delta m_{41}^2=0$ eV$^2$. Here only one NF far detector at 4000~km is used (without near detectors) in the left panel, whereas in the right panel two long-baseline detectors at 4000~km and 7500~km together with the near detectors at 2~km are included. The solid curves assume $\theta_{13}=0^\circ$, while the dashed curves assume $\theta_{13}=12^\circ$. The best-fit values are marked in the figure, the one of $\theta_{24}$ is zero. The MINOS curves are taken from \Ref~\cite{Adamson:2010wi}.}
\end{figure}

As similar kind of analysis can be performed in the special case $\vldm \rightarrow 0$, for which also no additional mass squared difference appears and $m_1 = m_4$. In this case, we have some sensitivity to $\theta_{34}$ in $\mathcal P_{\mu \mu}$, as shown in \App~\ref{pmm3}.  The comparison between MINOS and the Neutrino Factory is shown for this special case in \figu{minos-mod1}, and again the Neutrino Factory has an excellent sensitivity. The marginalization over $\vldm$ and the other parameters will, however, destroy the sensitivity to $\theta_{34}$, since there is no sensitivity in \figu{theta-mass} for $\vldm \rightarrow 0$. This is another good example that the set of assumptions determines the outcome, whereas the general analysis would simply produce no sensitivity.

\subsubsection*{Comparison to two-flavor limits}

Especially for short-baseline new physics searches, in the literature often two baseline limits are discussed. That is, at a short baseline where no standard oscillations occur, sterile neutrinos are searched for using a two flavor fit. An example is the short-baseline electron neutrino disappearance considered in the two flavor limit:
\begin{equation}
P_{ee} = 1 - \sin^2 (2 \theta ) \, \sin^2 \left( \frac{\Delta m^2 L}{4 E} \right) \, .
\label{equ:sbl}
\end{equation}
In \Ref~\cite{Giunti:2009en}, even the test of CPT invariance at such a short baseline was considered. From our discussion in \Sec~\ref{sec:ana}, it is clear that the short-baseline disappearance channels are the key measurements for the active-sterile-mixings, because they are uniquely sensitive to one parameter (or one mixing matrix element). On the other hand, the appearance channels, such as the tau neutrino appearance in NOMAD and CHORUS or the electron neutrino appearance in MiniBooNE, measure combinations of parameters. A signal in this channel would inevitably point towards new physics, but these channels are not very useful to limit individual active-sterile mixing parameters (or mixing matrix elements).

\equ{sbl} can be directly related to our parametrization with the identification $\theta \rightarrow \theta_{14}$ and $\Delta m^2 \rightarrow \vldm$. Correspondingly, the results of our analysis for $\theta_{14}$ are similar to \Ref~\cite{Giunti:2009en} (see Fig.~3). On the other hand, it was emphasized in \Ref~\cite{Giunti:2009en} that cross section measurements will spoil the disappearance measurements, and that these can be controlled with additional near detectors very close to the source. Since there is no CP violation in short-baseline disappearance measurements (neither from CP phases nor from the MSW effect), the disappearance channels can also be used for clean CPT invariance tests.

Note that a CPT invariance test of the MiniBooNE result would be, in principle, also possible at the Neutrino Factory, because both $P_{\bar \mu \bar e}$ and $P_{\mu e}$ could be tested. However, for these appearance channels electron charge identification is required to distinguish, for example, $P_{\bar \mu \bar e}$ from the $P_{ee}$ background. Therefore, the direct test of the recent anomaly~\cite{AguilarArevalo:2010wv} may be difficult.

\section{Summary and conclusions}
\label{sec:summary}

In this work, we have discussed sterile neutrinos beyond LSND, \ie, sterile neutrinos with small active-sterile mixings and an arbitrary $\vldm$. We have used the Neutrino Factory for the simulation, since this experiment can be used for a self-consistent approach including near and far detectors. We have used the simplest possible hypothesis, namely one extra sterile neutrino in a 3+1-like scheme in which the standard scenario is recovered for small active-sterile mixing angles. While we obtain the expected sensitivity to the active-sterile mixing in the large  LSND-motivated $|\vldm| \gtrsim 1\, \mathrm{eV}^2$ region, we also find sensitivity close to the atmospheric $\ldm$. We have pointed out that there are no global fits for this case yet, and a re-analysis of atmospheric and solar data should demonstrate what we can learn for $\vldm \ll 1\, \mathrm{eV}^2$. Note that 
recent cosmological fits point towards one or two such light sterile neutrinos~\cite{Hamann:2010bk}.

We have demonstrated that, especially at the short baselines, the disappearance channels are the primary channels of interest for light sterile neutrino constraints, no matter if a parametrization-independent approach or a particular parametrization for the mixing matrix is used. For the combined analysis of short and long baselines including charged and neutral current matter effects, however, we have used a particular parametrization. We have demonstrated that the most general constraint on sterile species can be shown as exclusion limits in the $\theta_{i4}$--$\vldm$--planes ($i=1$, $2$, $3$), similar to the CHOOZ limit for $\theta_{13}$--$\ldm$. The Neutrino Factory turns out to have excellent sensitivity to the three mixing angles in a wide range of $\vldm$. However, one of the three mixing angles ($\theta_{14}$ in our parametrization) can only be very well measured for large $|\vldm| \gtrsim 1\, \mathrm{eV}^2 $ at the near detectors, and one ($\theta_{34}$ in our parametrization) better for small $|\vldm| \sim  |\ldm|$, whereas the third ($\theta_{24}$) can be measured in the combined range. An electron neutrino disappearance channel at the long baselines could solve this problem  for $\theta_{14}$, which is, however, difficult at the Neutrino Factory, because electron charge identification might be required. An improved measurement of $\theta_{34}$ for large $|\vldm| \gtrsim 1\, \mathrm{eV}^2 $ would require a hypothetical $\nu_\tau$ disappearance channel. We have also investigated the impact of different mass orderings (\cf, \figu{MH}) on the sensitivities. We have found that there is a qualitative difference between the cases $\ldm \sim \vldm$ and $\ldm \sim -\vldm$ due to different matter effects. Furthermore, we have tested the impact of additional $\nu_\tau$ detectors with an aggressive 10~kt OPERA-like detector at the short and long baselines, and we have not found any significant effect on the sensitivities.

Apart from the general constraints, we have compared our analysis to special cases in the literature. For instance, we have tested the case $|\vldm| \gtrsim 1\, \mathrm{eV}^2 $ leading to averaging of the fast oscillations at the long baselines. We have looked into the special case of a  $\theta_{24}$-$\theta_{34}$ fit.
We have shown that this assumption involves subtleties, especially if additional near detectors are considered, because the assumption for $\vldm$ implies that there may be effects in the near detectors. We have illustrated that in this case additional assumptions are required, and the interpretation of the sensitivities becomes strongly assumption-dependent.  This does not apply to our general bounds we discussed above. We have also tested the impact of $\nu_\tau$ detection (``discovery channel''), and we could find a marginal improvement. The main sensitivity, however, comes from the muon neutrino disappearance channel for $E_\mu=25 \, \mathrm{GeV}$. We have moreover compared the Neutrino Factory to a recent MINOS analysis using the same assumptions, and we have found excellent sensitivities. However, including the full marginalization over the unknown parameters, the sensitivities were basically destroyed.

In conclusion, although the sensitivities of the Neutrino Factory to light sterile neutrinos in the atmospheric mass squared range may  not be extremely compelling, these bounds are very robust and independent of any special assumptions. If special assumptions are used, such as often done in the literature, the sensitivities look extremely good. Of course our results are not the most general case, one could also study $3+N$ scenarios. However, even for the $3+1$ scenario, this analysis is extremely challenging because there can be interference between the known and the unknown mass squared differences. A re-analysis of atmospheric and solar experiments could provide clues on what one can learn already from existing experiments. Finally, even the test in supernovae may be interesting, because one may expect additional resonances or new (neutral current) matter effects. 

\subsubsection*{Acknowledgments}

We would like to thank Carlo Giunti and Marco Laveder for lively discussions during the first stages of this work, and Michele Maltoni for useful discussions.
We would also like to thank Xiaobo Huang, Jiajie Ling, Anthony Mann, Brian Rebel, and Alexandre Sousa in the MINOS collaboration for help to make a direct comparison between their results and ours possible.

This work has been supported by the
Emmy Noether program of Deutsche Forschungsgemeinschaft (DFG), contract
no. WI 2639, by the DFG-funded Graduiertenkolleg 1147 ``Theoretical astrophysics and particle physics'' [J.T.], and by the European Union under the European Commission
Framework Programme~07 Design Study EUROnu, Project 212372.

\appendix
\section{$\boldsymbol{P_{\mu\mu}}$ in the limit $\boldsymbol{|\Delta_{41}| \ll \Delta_{31}}$}
\label{pmm3}

In this appendix we show the transition probability $P_{\mu\mu}$ in the limit $|\Delta_{41}| \ll \Delta_{31}$ and $\Delta_{21}=0$.
We study the explicit dependence on $\theta_{34}$ to all orders in perturbation theory, while keeping $\theta_{23}=45^\circ$ and all
other mixing angles vanishing. We get:
\bea
P_{\mu\mu}&=& \frac{C^2(1+c_{34}^4) + (-1+c_{34}^4)\Delta^2_{31} \sin^2 C}{C^2\,(1+c_{34}^2)^2}+ \nn \\ 
&&  \label{first} \\ \nn &&\frac{2 c_{34}^2 \sin C (\Delta_n \sin \Delta_{n}-c_{34}^2 \Delta_{31}\sin \Delta_{31})}{C\,(1+c_{34}^2)^2} \eea
where 
\bea
C&=&\sqrt{2  c_{34}^2  \Delta_{31} \Delta_n + \Delta_{31} ^2+ \Delta_n^2}\,.
\eea
We see that the  dependence on $\theta_{34}$ (through the $c_{34}$ terms) appears in terms with different energy dependencies, which 
makes a spectral analysis at the Neutrino Factory useful to get a non-vanishing sensitivity even in  the 
$|\Delta_{41}| \ll \Delta_{31}$ limit. Interestingly enough, the matter effects are driven by the neutral current potential 
$\Delta_n$, because of the vanishing $\theta_{13}$ approximation.

% \bibliographystyle{apsrev}
% \bibliography{references}

\end{document}